\documentclass[]{jfm} 

\usepackage{graphicx}
\usepackage{newtxtext}
\usepackage{newtxmath}
\usepackage{natbib}
\usepackage{hyperref}
\hypersetup{
    colorlinks = true,
    urlcolor   = blue,
    citecolor  = black,
}

\newcommand{\RomanNumeralCaps}[1]

\title{Prolate spheroids settling in a quiescent fluid: clustering, microstructures and collisions}

\author{Xinyu Jiang\aff{1}, Chunxiao Xu\aff{1}, Lihao Zhao\aff{1}\corresp{\email{zhaolihao@tsinghua.edu.cn}}}

\affiliation{\aff{1}AML, Department of Engineering Mechanics, Tsinghua University, Beijing 100084, China}

\begin{document}
\maketitle

\begin{abstract}
In this study, we investigate the sedimentation of prolate spheroids in a quiescent fluid by means of the particle-resolved direct numerical simulation. With the increase of the particle volume fraction $\phi$ from $0.1\%$ to $10\%$, we observe a non-monotonic variation of the mean settling velocity of particles, $\langle V_s \rangle$. By virtue of the Voronoi analysis, we find that the degree of particle clustering is highest when $\langle V_s \rangle$ reaches the local maximum at $\phi=1\%$. Under the swarm effect, clustered particles are found to preferentially sample downward fluid flows in the wake regions, leading to the enhancement of the settling speed. As for lower or higher volume fraction, the tendency of particle clustering and the preferential sampling of downward flows are attenuated. Hindrance effect becomes predominant when the volume fraction exceeds 5\% and reduces $\langle V_s \rangle$ to less than the isolated settling velocity. Particle orientation plays a minor role in the mean settling velocity, although individual prolate particles still tend to settle faster in suspensions when they deviate more from the broad-side-on alignment. Moreover, we also demonstrate that particles are prone to form column-like microstructures in dilute suspensions under the effect of wake-induced hydrodynamic attractions. The radial distribution function is higher at a lower volume fraction. As a result, the collision rate scaled by the particle number density decreases with the increasing volume fraction. By contrast, as another contribution to the particle collision rate, the relative radial velocity for nearby particles shows a minor degree of variation due to the lubrication effect.
\end{abstract}

\section{Introduction} \label{sec:intro}
Particle-laden flows are commonly encountered in natural and industrial processes, such as the transport of pollution in the air or underwater, the marine snow generated by settling plankton or microplastics, and precipitation in the atmosphere \citep{pruppacherMicrophysicsCloudsPrecipitation2010,guazzelliFluctuationsInstabilitySedimentation2011,trudnowskaMarineSnowMorphology2021}.
One of the fundamental challenges in these applications is the gravity-driven sedimentation of particles in fluids, which involves complex interactions between moving particles and the carrying fluid flow \citep{guazzelliFluctuationsInstabilitySedimentation2011}, as well as collisions among the dispersed particles \citep{ayalaEffectsTurbulenceGeometric2008}. 

\subsection{Settling of spherical particles} \label{subsec:settle_sphere_intro}
In past decades, a series of experimental and numerical studies on the settling of an isolated sphere have been carried out. These earlier studies revealed that the dynamic mode and moving speed of a single falling/rising sphere are determined by two dimensionless parameters, the density ratio $\alpha$ and the Galileo number $Ga$ \citep{jennyInstabilitiesTransitionSphere2004,horowitzEffectReynoldsNumber2010,zhouChaoticStatesOrder2015,raaghavPathInstabilitiesFreely2022}. The former measures the inertia of the solid particle, and the later quantifies the ratio between the buoyancy and viscous force acting on the sphere. In this study, our focus is on settling particles so the density ratio is greater than unity. 
One can also describe this problem by defining a dependent parameter, the Reynolds number $Re_t$, based on the a-posteriori settling velocity $V_t$ and the diameter of the sphere $D$.
In the creeping flow regime ($Re_t \ll 1$), the sphere settles vertically with a constant settling velocity by balancing the buoyancy force with the Stokes drag. With the increase of the Reynolds number to a finite value, the introduction of fluid inertia breaks the fore-aft symmetry of the fluid flow around the settling sphere so the wake emerges. As $Re_t$ increases, the change of the rear-wake morphology can trigger the path instability of the settling sphere. A variety of settling modes, including vertical, oblique, zigzag, helical and chaotic motions, could be observed with varying inertia of the fluid and particle \citep{jennyInstabilitiesTransitionSphere2004,horowitzEffectReynoldsNumber2010,ernWakeInducedOscillatoryPaths2012,zhouChaoticStatesOrder2015,raaghavPathInstabilitiesFreely2022}.

Concerning the settling motion of a pair of particles, the hydrodynamic interaction between them must be taken into account. A typical phenomenon is the drafting-kissing-tumbling (DKT) process of a pair of initially vertical-aligned settling spheres in the inertial-flow regime \citep{fortesNonlinearMechanicsFluidization1987,glowinskiFictitiousDomainApproach2001}. Specifically speaking, in the first stage (drafting stage), the trailing particle accelerates its settling motion as it resides in the wake of the leading particle.
The two particles exhibit an attractive relative motion during this stage. Subsequently, the two particles touch and form an elongated body aligning along the vertical direction (kissing stage). 
However, settling with this configuration is unstable, so the particle pair tumbles and separates under the effect of hydrodynamic interaction (tumbling stage). In this stage, the two particles behave as if they repel each other, and the originally trailing particle becomes the leading one.
Hence, the DKT process reflects the complicated hydrodynamic interaction between a pair of settling particles. 

When considering the sedimentation of a group of particles, the most well-known phenomenon is the hindered settling motion of dispersed particles in suspensions \citep{richardsonSedimentationSuspensionUniform1954}, i.e. the reduction of the mean settling velocity of particles with the increase of the volume fraction $\phi$. The physical explanation of this hindrance effect is as follows. To maintain zero net flux of the whole flow system, a mean upward fluid flow is generated to counteract the downward flux of settling particles. As a result, the upward mean flow increases the hydrodynamic drag acting on the particle phase and thus reduces the mean settling velocity \citep{difeliceSedimentationVelocityDilute1999}. 
An empirical formula of the hindered settling velocity of particles in the creeping-flow regime was also proposed by \citet{richardsonSedimentationSuspensionUniform1954}. More recently, the hindered settling velocity of particles with finite fluid inertia was also observed in the numerical studies by means of the two-way coupling point-particle simulation \citep{climentNumericalSimulationsRandom2003} or the particle-resolved direct numerical simulation (PR-DNS) \citep{yinHinderedSettlingVelocity2007,zaidiDirectNumericalSimulation2014}. The empirical expression of the hindered settling velocity has also been improved in a series of studies to incorporate the finite-Reynolds-number correction \citep{garsideVelocityVoidageRelationshipsFluidization1977,difeliceSedimentationVelocityDilute1999,yinHinderedSettlingVelocity2007}.

However, over the past two decades, a striking enhancement of the mean particle settling velocity has been observed under certain conditions, thanks to the state-of-the-art PR-DNS of the particle sedimentation \citep{kajishimaInteractionParticleClusters2002, kajishimaInfluenceParticleRotation2004, uhlmannSedimentationDiluteSuspension2014, zaidiDirectNumericalSimulation2014}. 
Investigations into the particle spatial distribution have shown that the enhanced settling velocity is always associated with the formation of column-like particle clusters \citep{doychevDynamicsFinitesizeSettling,uhlmannSedimentationDiluteSuspension2014,zaidiDirectNumericalSimulation2014}. Later on, the experimental work conducted by \citet{huismanColumnarStructureFormation2016} also confirmed these numerical observations.
\citet{zaidiDirectNumericalSimulation2014} and \citet{moricheClusteringLowaspectratioOblate2023} attributed the formation of particle clusters to the DKT-like interactions among settling particles, but this phenomenon can only be observed in dilute suspensions when the Reynolds number $Re_t$ is sufficiently high. Conversely, at low Reynolds numbers, the weaker wake-induced attraction among particles results in orderly particle arrangements rather than particle clustering \citep{yinHinderedSettlingVelocity2007,zaidiDirectNumericalSimulation2014,zaidiDirectNumericalSimulations2015}. While, in dense suspensions, the short distances between particles disrupt particle wakes and thus inhibit the formation of particle clusters as well \citep{zaidiParticleVelocityDistributions2018}.
Readers can refer to \citet{chouippeResultsParticleresolvedSimulations2023} for the review of previous studies on this problem.

\subsection{Settling of non-spherical particles}
In practice, the shape of dispersed particles is commonly non-spherical. For instance, ice crystals in clouds, plankton in the marine, and dusts in the atmosphere are usually disk-like or rod-like in shape \citep{shawParticleTurbulenceInteractionsAtmospheric2003,malliosEffectsDustParticle2020,slomkaCollisionRodsQuiescent2020}. For simplicity, non-spherical particles are often modelled by smooth-surface prolate/oblate spheroids or by polyhedrons with edges. Compared to spherical particles, the orientational behavior and rotational motion of non-spherical particles add complexity to their dynamics in fluid flows \citep{rahmaniFreeFallingRising2014,vothAnisotropicParticlesTurbulence2017}.

As for a single spheroid settling in a quiescent fluid, the particle would maintain its initial orientation in the creeping flow owing to the vanishing hydrodynamic torque. Thus, the settling velocity of the spheroid is orientation-dependent in this regime \citep{happelLowReynoldsNumber1983}. However, when the fluid inertia is taken into account, a non-negligible hydrodynamic torque reorients the settling spheroid to a broad-side-on alignment \citep{khayatInertiaEffectsMotion1989,ardekaniNumericalStudySedimentation2016,dabadeEffectInertiaOrientation2016}. Additionally, when the fluid inertia is strong enough to trigger wake instability, the settling motion of the spheroid transitions from the steady vertical falling to complicated unsteady modes, similar to the case of a settling sphere. 
The velocity and mode (including spiral, zigzag/fluttering, tumbling, and chaotic) of the settling motion are jointly determined by the density ratio, Galileo number and the shape of the spheroid \citep{chrustNumericalSimulationDynamics2013,ardekaniNumericalStudySedimentation2016,zhouPathInstabilitiesOblate2017,moricheSingleOblateSpheroid2021}, and can also be altered by the presence of walls \citep{huangSedimentationEllipsoidalParticle2014,yangSedimentationOblateEllipsoid2015}.

Moreover, according to the numerical work by \citet{ardekaniNumericalStudySedimentation2016}, the DKT process between a pair of settling spheroids is quite different from that of spherical particles. 
As for a pair of oblate particles with an aspect ratio (the ratio between the polar and equator radius) $\lambda=1/3$, the two particles do not undergo the tumbling stage after they approach and touch. Instead, they fall with a steady pilled-up configuration, as if they are stuck together \citep{ardekaniNumericalStudySedimentation2016}. 
Regarding a pair of prolate spheroids with $\lambda=3$, the DKT process is more complicated and dependent on their initial relative angle \citep{ardekaniNumericalStudySedimentation2016}. If the symmetry axes of the two prolate particles are initially parallel, the DKT process is similar to that of spherical particles. While, if the symmetry axes are perpendicular at the beginning, a stable cross-like configuration is formed and the two prolate particles do not separate for a long time after they touch, similar to the case of oblate particles. In principle, the attraction zone (within which the trailing particle can be attracted by the leading one) is larger, and the interaction time (the time duration for the particle pair to keep in touch) is longer for the DKT process of spheroidal particle pairs, compared to that of spherical ones \citep{ardekaniNumericalStudySedimentation2016,moricheClusteringLowaspectratioOblate2023}. To conclude, the particle shape plays an important role in the hydrodynamic interaction between settling particles.

As regards the settling of a large number of non-spherical particles, an increased settling velocity of elongated fibres was observed in the creeping flow regime due to the formation of particle streamers aligning in the gravitational direction \citep{kuuselaCollectiveEffectsSettling2003,saintillanGrowthConcentrationFluctuations2006,shinStructureDynamicsDilute2009}. 
In the finite-fluid-inertia regime, \citet{seyed-ahmadiSedimentationInertialMonodisperse2021} numerically studied the settling motion of cubic particles. In contrast to the clustered settling spheres at $Ga=160$ and $\phi=1\%$, the spatial distribution of settling cubes is closer to a random distribution in the same parameter setup, which was attributed to the greater rotational rate of settling cubes. \citet{fornariClusteringIncreasedSettling2018} simulated the sedimentation of oblate spheroids with $\lambda=1/3$ and $Ga=60$ at different particle volume fractions. 
They reported appreciable particle clustering for settling oblate particles at a relatively low Reynolds number ($Re_t=38.7$), and considerable enhancement of the mean settling velocity up to $\langle V_s \rangle \approx 1.33 V_t$ at $\phi=0.5\%$.
Similar results were also reported in the recent work by \citet{moricheClusteringLowaspectratioOblate2023}, who considered the low-aspect-ratio oblate spheroids with $\lambda=2/3$ and higher Galileo number with $Ga=111$ and 152. 
As for the case of prolate particles, \citet{luDynamicsSuspensionsProlate2023} simulated the settling of prolate spheroids with $\lambda=2$ and $Ga=41.8$ at $\phi=2.2\%, 5.5\%$ and $9.9\%$ using a relatively small periodic computational domain. They reported a decreased mean particle settling velocity and a transition from the hydrodynamic-interaction-dominated regime to the particle-collision-dominated regime with the increase of $\phi$. 

\subsection{Particle collisions}
The collision rate among dispersed particles plays an important role in the particle coagulation in fluid flows, which are relevant to many industrial and natural processes.
In the past, a plenty of work has been carried out to study the collision and coagulation of point-like spherical particles in turbulent flows in the framework of one-way coupling approach \citep{saffmanCollisionDropsTurbulent1956,sundaramCollisionStatisticsIsotropic1997,wangStatisticalMechanicalDescription2000,ayalaEffectsTurbulenceGeometric2008}. Readers can refer to the reviews by \citet{grabowskiGrowthCloudDroplets2013} and \citet{pumirCollisionalAggregationDue2016} for more details. Recently, some researchers extended the work to non-spherical particles, and demonstrated that the orienational behavior of elongated or flattened particles enhances their collision rate in turbulence \citep{juchaSettlingCollisionSmall2018,siewertCollisionRatesSmall2014,slomkaCollisionRodsQuiescent2020,arguedas-leivaElongationEnhancesEncounter2022,grujicCollisionsElongatedSettling2024}.
However, when further considering the intricate particle-fluid and particle-particle interactions, the understanding of particle collision rates remains limited. In \citet{wangTheoreticalFormulationCollision2005}, the hydrodynamic interactions among particles were addressed by adding the particle-induced disturbance into the background turbulence. These disturbance can either augment or attenuate collision rate, depending on whether they act as the far-field or near-field influence. In the framework of PR-DNS, \citet{chenParticleresolvedDirectNumerical2020} simulated the transport of spherical particles in the homogeneous isotropic turbulence, and studied the collision rate of bidispersed inertial particles. Furthermore, \citet{fornariSettlingFinitesizeParticles2019} simulated settling spherical particles in both the quiescent fluid and the turbulent environment, and examined the effect of particle spatial distribution and relative motion on the collision rate. 
However, to the best of our knowledge, the collision rate of settling non-spherical particles with the full consideration of fluid-particle and particle-particle interactions has not been investigated so far.

\subsection{Objective of the present study}
According to the above literature review, we are still far from achieving a comprehensive understanding of settling non-spherical particles in suspensions. In the present work, we investigate the sedimentation of prolate particles in an initially quiescent fluid by means of PR-DNS. In particular, considering the significant impact of hydrodynamic interactions on the dynamics of pairwise settling prolate particles \citep{ardekaniNumericalStudySedimentation2016}, we aim to explore the collective behavior of settling prolate particles at varying volume fractions. 
This work is motivated by two concerns. First, although the enhancement of particle settling velocity and particle clustering down to $\phi \approx 0.02\%$ have been reported for spherical particles \citep{doychevDynamicsFinitesizeSettling,huismanColumnarStructureFormation2016}, to the best of the authors' knowledge, little is known about the sedimentation of non-spherical particles with $\phi<0.5\%$ in the inertial-flow regime. Hence, we study the settling motion of prolate spheroids within a wide range of the volume fraction from $\phi=0.1\%$ to $10\%$, with the particle aspect ratio and Galileo number fixed at $\lambda=3$ and $Ga=80$,
following the study of a single and a pair of settling prolate particles by \citet{ardekaniNumericalStudySedimentation2016}. The particle-fluid density ratio is set as $\alpha=2$, which is a typical value for a solid-liquid system \citep{seyed-ahmadiSedimentationInertialMonodisperse2021}.
Interestingly, we observe a non-monotonic variation of the mean particle settling velocity as $\phi$ increases, so we further investigate the influences of particle clustering, hindrance effect and particle orientation on the settling speed of prolate spheroids. Second, the collision rate among settling non-spherical particles with finite sizes is not well understood so far. 
Therefore, we also investigate on this issue and scrutinize the particle pair statistics that are essential in determining the particle collision rate.

The remainder of this paper is organized as follows. In section \ref{sec:setup}, we describe the physical problem and the simulation set-ups of this study. Then, in section \ref{sec:result}, we analyze the statistics of particle motions and spatial distributions, followed by the examination of the collision rate of dispersed particles at different particle volume fractions. Finally, we summarize the findings and draw conclusions in section \ref{sec:conclud}.

\section{Simulation set-ups}\label{sec:setup}
\begin{figure}
    \centering
    \includegraphics[width=0.5\textwidth]{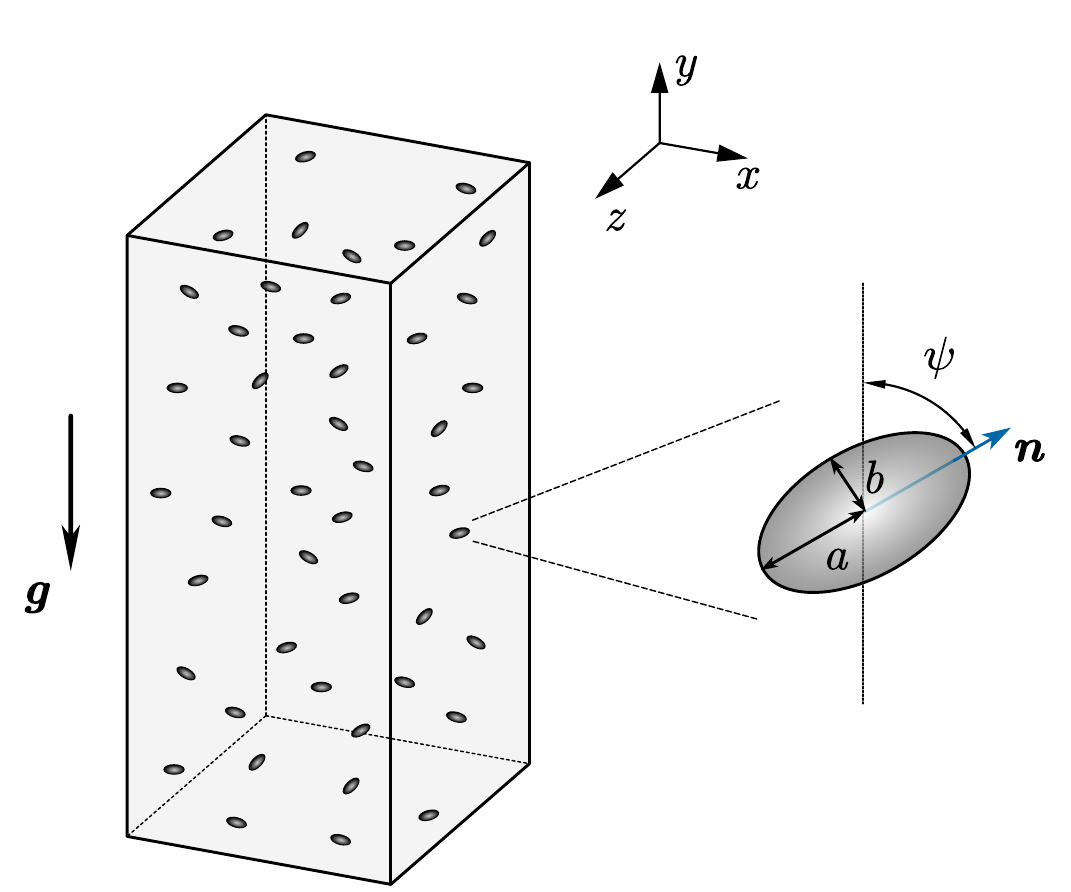}
    \caption{Schematic representation of settling prolate particles in a quiescent fluid. The semi-major and semi-minor axes of the prolate particle have a length of $a$ and $b$, respectively. The unit vector along the symmetry axis of the prolate particle is denoted by $\boldsymbol{n}$. The angle between the vector $\boldsymbol{n}$ and the positive $y$ direction is defined as the pitch angle $\psi$. The gravity is applied in the negative $y$ direction with an acceleration of $\boldsymbol{g}$.}
    \label{fig:schematic}
\end{figure}

\begin{table}
    \begin{center}
    \def~{\hphantom{0}}
    \begin{tabular}{ccccc}
    Case & $\phi$  & $\left[L_x \times L_y \times L_z\right]/D_{eq}^3$ & $N_p$ & $N_{cell}$ \\
      1  & $0.1\%$ & $32\times 100 \times 32$                 & 196  & 1.416 billion \\
      2  & $0.5\%$ & $32\times 100 \times 32$                 & 978  & 1.416 billion \\
      3  & $1\%$   & $32 \times 100 \times 32$                & 1956 & 1.416 billion \\
      4  & $2\%$   & $32 \times 100 \times 32$                & 3911 & 1.416 billion \\
      5  & $5\%$   & $ 24 \times 60 \times 24$                & 3300 & 0.478 billion\\
      6  & $10\%$  & $20 \times 50 \times 20$                 & 3820 & 0.276 billion\\ 
    \end{tabular}
    \caption{Simulation set-ups for settling prolate particles with different particle volume fraction. The total number of grid cells used for the fluid flow simulation is denoted by $N_{cell}$.}
    \label{tab:config}
    \end{center}
\end{table}

The configuration of simulations in the present study is sketched in figure \ref{fig:schematic}. The prolate particle with an aspect ratio $\lambda=a/b=3$ is considered in this study. The Galileo number, defined by $Ga=\sqrt{(\alpha-1)|\boldsymbol{g}|D_{eq}^3}/\nu$, is set as $Ga=80$. Here, $D_{eq}=2(a b^2)^{1/3}$ is the equivalent diameter (defined as the diameter of a sphere with the same volume of the prolate spheroid), $\alpha$ is the particle-fluid density ratio which is set as $\alpha=2$, and $\boldsymbol{g}$ is the acceleration induced by gravity. 
Under this parameter set-up, a single prolate spheroid settles vertically with the broad-side-on orientation (see appendix \ref{appsec:isolated}). The settling Reynolds number corresponding to the isolated settling velocity $V_t$ is $Re_t=V_t D_{eq} /\nu = 61.8$.

To study the effect of volume fraction on the settling motion of prolate particles, we consider six simulation cases as listed in table \ref{tab:config}. The particle volume fraction $\phi$ is defined by $\phi=(\pi N_p D_{eq}^3)/(6 L_x L_y L_z)$, in which $N_p$ denotes the number of particles, and $L_x$, $L_y$ and $L_z$ represent the length of the computational domain in the $x$, $y$ and $z$ directions, respectively. The periodic boundary condition is imposed in each direction of the computational domain. The grid resolution is set as $\Delta h= D_{eq}/24$ in the simulations, which is fine enough to fully resolved the particle-fluid interactions (see appendix \ref{appsec:isolated}). The time step used in the present simulations is $\Delta t=0.01D_{eq}/V_t$, which corresponds to a Courant number of $CFL=0.24$ for the adopted grid resolution based on the settling velocity of an isolated particle.
The size of the computational domain and the total number of grid cells are provided in table \ref{tab:config}. As the gravity is applied in the negative $y$ direction, the computational domain along this vertical direction is set longer than the other two lateral directions. 
Note that we reduce the size of the computational domain for the cases with $\phi \ge 5\%$ to save the computational cost. This is reasonable because the decorrelation of the fluid velocity is more rapid as the particle volume fraction increases \citep{zaidiDirectNumericalSimulation2014,zaidiParticleVelocityDistributions2018}. 
We have checked that the two-point correlation functions for the fluid velocity fluctuations decay to less than 0.3 at the longest distance, except for the vertical velocity fluctuations along the vertical direction at $\phi=0.5\%$ and $1\%$. The slow decorrelation at $\phi=0.5\%$ and $1\%$ is attributed to the column-like particle clustering in these two cases (see section \ref{subsubsec:cluster} for more details), which was also reported in \citet{uhlmannSedimentationDiluteSuspension2014} and \citet{moricheClusteringLowaspectratioOblate2023}. However, according to \citet{zaidiDomainSizeEffects2021}, the statistics of the particle dynamics are not affected by the size of the computational domain when it is larger than 10 times of the particle size. 
Hence, the present computational domain is sufficiently large for obtaining qualitatively reliable results and we do not enlarge the domain size considering the affordability of the computational cost.

As for the initial configuration of the dispersed particles, we adopt the method proposed by \citet{anoukouRandomDistributionPolydisperse2018} to generate non-overlap prolate particles with random spatial distribution and random orientations in the simulation. Released from rest, particles accelerate their settling motion under the action of gravity, and the flow system would eventually reach a statistically steady state after a developing transient.
The statistics presented in section \ref{sec:result} are collected in the steady state. Specifically, the data within a time window of $200D_{eq}/V_t$ are used for computing the statistics for most cases, except for the extension of this time window to $350D_{eq}/V_t$ in the most dilute case with $\phi=0.1\%$ because of the considerably reduced number of particles.

To realize the PR-DNS of the present particle-laden flow system, we use the immersed boundary method (IBM) to resolve the particle-fluid interactions \citep{peskinImmersedBoundaryMethod2002, iaccarinoImmersedBoundaryMethod2004}. In particular, the fluid flow is simulated by numerically solving the incompressible Navier-Stokes equations with a second-order finite difference method \citep{kimImplicitVelocityDecoupling2002}. The six-degree-of-free motion of the dispersed particles are simulated by integrating the Newton-Euler equations.
Additionally, we employ the direct-forcing immersed boundary method \citep{uhlmannImmersedBoundaryMethod2005,breugemSecondorderAccurateImmersed2012} for the coupling between the particle motion and the fluid flow. 
Moreover, to model the inter-particle collisions, a soft-sphere collision model together with a lubrication correction is employed \citep{costaCollisionModelFully2015,ardekaniNumericalStudySedimentation2016}.
More details about the computational method adopted in the present study are provided in Appendix \ref{app:method}.

\section{Results and discussion} \label{sec:result}
\subsection{Particle settling velocity} \label{subsec:stats}
\begin{figure}
    \centering
    \includegraphics[width=0.45\textwidth]{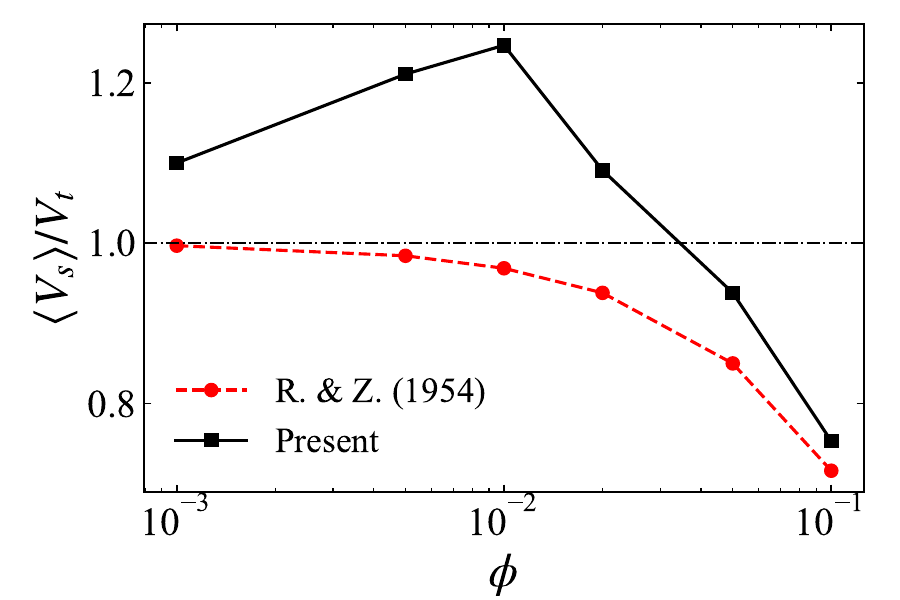}
    \caption{Mean settling velocity of dispersed particles at different volume fraction. The empirical correlation of the hindered settling velocity \citep{richardsonSedimentationSuspensionUniform1954} (depicted by the red dashed line) is included for comparison.}
    \label{fig:vel_sett_mean}
\end{figure}
The first observable we discuss is the mean settling velocity of dispersed particles. Here, the settling velocity is defined as the component of particle velocity along the gravitational direction, i.e. $V_s=\boldsymbol{v}\cdot \boldsymbol{e}_g=-v_y$.
As shown in figure \ref{fig:vel_sett_mean}, the mean settling velocity $\langle V_s \rangle$ exhibits a non-monotonic variation with the increase of particle volume fraction from $\phi=0.1\%$ to $\phi=10\%$. Specifically, the mean settling velocity is greater than the settling velocity of an isolated particle when $\phi \le 2\%$, with a peak value of $\langle V_s \rangle \approx 1.25 V_t$ at $\phi=1\%$, and decreases to less than $V_t$ when the particle volume fraction exceeds $5\%$. 
In the following, we look into the particle clustering, hindrance effect, and the particle orientation to interpret the non-monotonic variation of the particle mean settling velocity with the change of the volume fraction.

\subsubsection{Particle clustering} \label{subsubsec:cluster}
In previous studies, it has been reported that the enhancement of the particle mean settling velocity is highly related to the formation of particle clustering \citep{kajishimaInfluenceParticleRotation2004,uhlmannSedimentationDiluteSuspension2014,fornariClusteringIncreasedSettling2018}. Thus, we first examine the spatial distribution of dispersed particles in the present simulations using the Voronoi analysis \citep{monchauxPreferentialConcentrationHeavy2010}. In this analysis method, the entire computational domain is partitioned into $N_p$ cells, i.e. Voronoi tessellations. The partitioning rule ensures that a given spatial point inside the $i$-th tessellation is closest to the centroid of the $i$-th particle among all particles. Accordingly, the spatial distribution of dispersed particles can be quantified by the statistics of the normalized volume of Voronoi tessellations, $\overline{V}_{Voro}(i)={V}_{Voro}(i) N_p/V_{tot}$, where ${V}_{Voro}(i)$ is the volume of the $i$-th Voronoi tessellation, and $V_{tot}$ the volume of the whole domain. 
If particles are orderly distributed in the space (like molecules in a crystal), the entire domain would be evenly partitioned so that $\overline{V}_{Voro}\equiv1$ and the standard deviation is $\sigma(\overline{V}_{Voro})=0$.
In contrast, in a system where particle clustering arises, the prevalence of particle accumulations (represented by small values of $\overline{V}_{Voro}$) and voids (represented by large values of $\overline{V}_{Voro}$) would increase the intermittency of the probability density function (p.d.f.) of and the standard deviation of $\overline{V}_{Voro}$ \citep{monchauxPreferentialConcentrationHeavy2010}. 

To quantify the degree of particle clustering, \citet{tagawaClusteringMorphologyFreely2013} defined the clustering indicator $C$ based on the standard deviation of the normalized Voronoi volume of dispersed particles as
\begin{equation} \label{eq:cluster_indicator}
    C = \sigma \left( {{{\overline V }_{Voro}}} \right)/\sigma \left( {{{\overline V }_{Voro,rand}}} \right),
\end{equation}
where the subscript `$rand$' represents the assembly of particles with a random spatial distribution. According to this definition, particles are considered to form clusters when the clustering indicator exceeds unity, and a higher level of clustering is identified by a greater value of $C$. 
As for point-like particles, the statistics of $\overline{V}_{Voro,rand}$ conforms to a Gamma distribution, yielding a standard deviation of $\sigma(\overline{V}_{Voro,rand})=0.447$ \citep{ferencSizeDistributionPoisson2007}. However, the value of $\sigma(\overline{V}_{Voro,rand})$ is a decreasing function of $\phi$ for non-overlapping finite-size particles \citep{uhlmannVoronoiTessellationAnalysis2020}. In the present work, we generate randomly distributed prolate particles with random orientations following the method proposed by \citet{anoukouRandomDistributionPolydisperse2018} and compute $\sigma(\overline{V}_{Voro,rand})$ accordingly. To ensure the convergence of $\sigma(\overline{V}_{Voro,rand})$ with sufficient samples, we repeat the generating process 1000 times for $\phi=0.1\%$, 200 times for $\phi=0.5\%$ and 100 times for other volume fractions.

\begin{figure}
    \centering
    \includegraphics[width=0.8\textwidth]{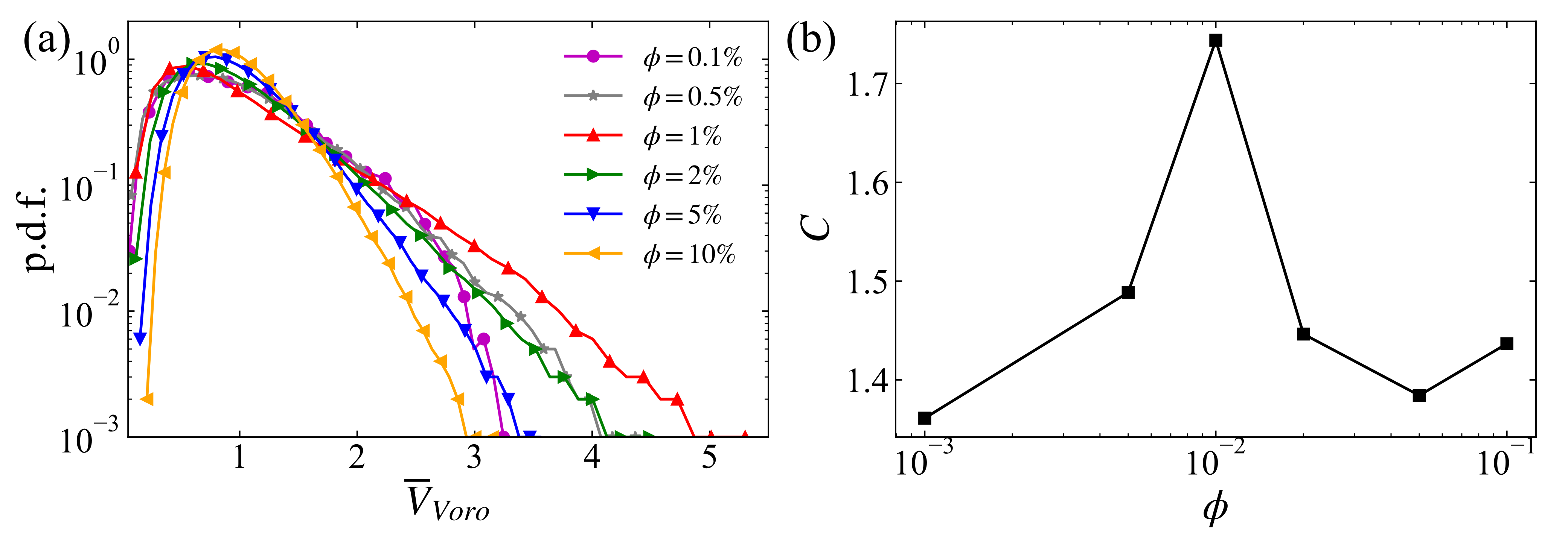}
    \caption{Results of the Voronoi analysis at different volume fraction. (a) P.d.f. of the normalized volume of Voronoi tessellations. (b) Clustering indicator $C$ at different volume fraction $\phi$.}
    \label{fig:Voronoi}
\end{figure}

The statistical distributions of the normalized Voronoi volume in the current work are illustrated in figure \ref{fig:Voronoi}(a). We can clearly observe {a raised tail} for the p.d.f. of $\overline{V}_{Voro}$ in the cases with $0.5\% \le \phi \le 2\%$. In contrast, the distribution of $\overline{V}_{Voro}$ is narrowed in the densest suspension at $\phi=10\%$. 
Furthermore, the variation of the clustering indicator $C$ as the function of the volume fraction is illustrated in figure \ref{fig:Voronoi}(b). 
Interestingly, the value of $C$ varies non-monotonically with a peak at $\phi=1\%$, which coincides with the highest mean settling velocity of particles as is observed in figure \ref{fig:vel_sett_mean}. 
For the cases with lower or higher volume fraction, the clustering indicator is reduced, although its value remains to be greater than unity. Thus, the particle clustering becomes less pronounced in more dilute or denser suspensions.

In the previous studies of particle sedimentation, the DKT-like interactions among settling particles are regarded as the essential mechanism in the formation of particle clusters \citep{kajishimaInfluenceParticleRotation2004,zaidiDirectNumericalSimulation2014,fornariClusteringIncreasedSettling2018, moricheClusteringLowaspectratioOblate2023}. In particular, during the drafting stage of a DKT event, a pair of particles can attract each other, reducing the distance between them. Furthermore, if these interacting particles attract additional particles before they separate, the number of accumulated particles can increase progressively, which eventually results in particle clustering in the suspension \citep{moricheClusteringLowaspectratioOblate2023}.
Compared to settling spheres, spheroidal particles are more likely to be drawn into the wake of a leading particle and tend to have a longer interaction time in the DKT process \citep{ardekaniNumericalStudySedimentation2016}. This explains the occurrence of clustered prolate particles in this study, similar to the behavior of settling oblate particles \citep{fornariClusteringIncreasedSettling2018, moricheClusteringLowaspectratioOblate2023}, in contrast to the absence of particle clustering of settling spheres at a comparable Reynolds number \citep{zaidiDirectNumericalSimulation2014}.

\begin{figure}
    \centering
    \includegraphics[width=0.99\textwidth]{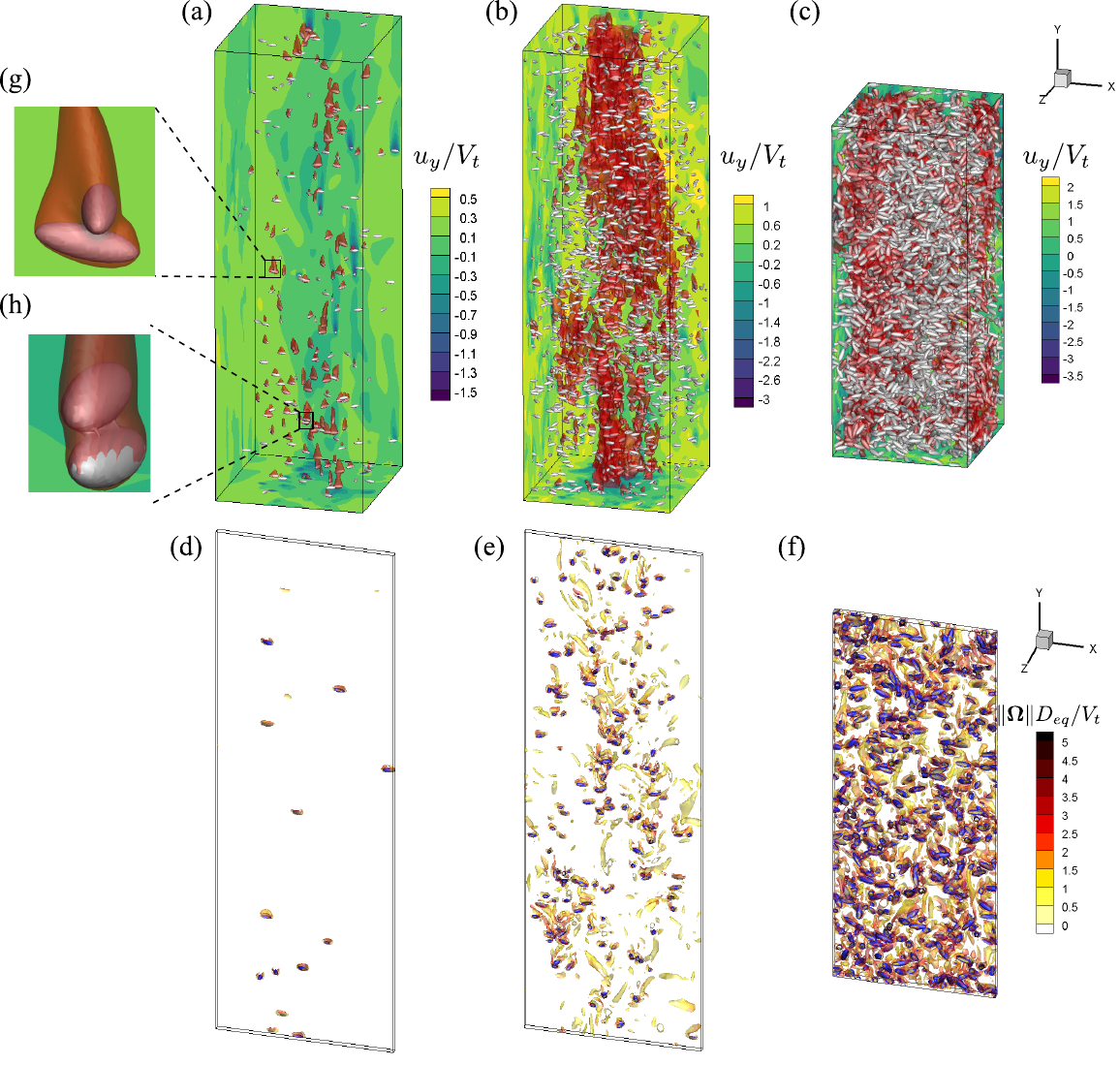}
    \caption{(a-c) Snapshots of the instantaneous flow field at (a) $\phi=0.1\%$, (b) $\phi=1\%$ and (c) $\phi=10\%$. Dispersed prolate particles are depicted in grey. The background contour represents  the vertical fluid vleoicty $u_y$ (normalized by $V_t$), with iso-surfaces of $u_y=-\langle V_s \rangle$ shown in red. (d-f) Vertical sections with a thickness of $D_{eq}$ along the $z$-direction taken from panels (a-c). Dispersed prolate particles are shown in blue. Iso-surfaces of the Q-criterion at $Q=0.2V_t^2/D_{eq}^2$ are colored by the magnitude of vorticity $\|\boldsymbol{\Omega}\|$ (normalized by $V_t/D_{eq}$). (g,h) Zoom-in views of panel (a) to illustrate the touching particle pairs at $\phi=0.1\%$.}
    \label{fig:instant}
\end{figure}

In figure \ref{fig:instant}, we provide the visualization of the flow system at three typical volume fractions $\phi=0.1\%$, $1\%$ and $10\%$.
In the most dilute case with $\phi=0.1\%$ (figure \ref{fig:instant}(a,d)), the particles are sparsely distributed in the space without appreciable particle clustering.
However, individual DKT events induced by the wake-related hydrodynamic interactions can still be found (see figure \ref{fig:instant}(g,h)). As for the case with $\phi=1\%$, while, we can evidently observe locally-accumulated particles and some regions devoid of particles (see figure \ref{fig:instant}(e)). Meanwhile, large-scale flow structures formed by the interconnected particle wakes are also illustrated in figure \ref{fig:instant}(b). These structures exhibit the footprint of the column-like particle clusters meandering along the vertical direction (with more details provided in section \ref{subsec:micro}). 
While, in another limit of dense suspension, particles are crowded in the space as shown in figure \ref{fig:instant}(c,f). The too small distance between neighboring particles frequently perturbs particle wakes, so as to inhibit the formation of particle clustering \citep{zaidiDirectNumericalSimulation2014}. Incidentally, the slight growth of the clustering indicator $C$ from $\phi=5\%$ to $10\%$ (see figure \ref{fig:Voronoi}(b)) may be related to the more ordered arrangement for the random-distributed particles, which reduces $\sigma(\overline{V}_{Voro,rand})$ in equation \eqref{eq:cluster_indicator}.

\begin{figure}
    \centering
    \includegraphics[width=0.8\textwidth]{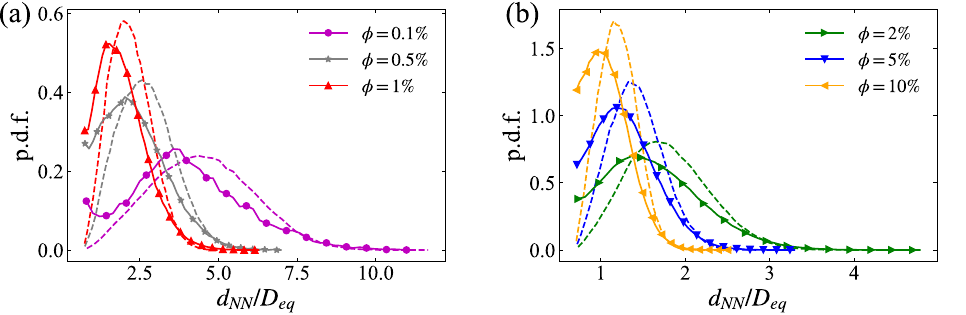}
    \caption{P.d.f. of the NND of particles at volume fraction (a) $\phi=0.1\%,\ 0.5\%,\ 1\%$ and (b) $\phi=2\%,\ 5\%,\ 10\%$. The solid lines are the results of the present simulations while the dashed lines represent the results of randomly distributed prolate spheroids with random orientation. Each line starts from $d_{NN}=2b$, which is the smallest center-to-center distance between two finite-sized prolate particles.}
    \label{fig:NND_compare}
\end{figure}

Moreover, we also look into the statistics of the nearest neighbor distance (NND) of particles in the present simulations. Here, the nearest neighbor distance of the $i$-th particle, denoted by $d_{NN}(i)$, is defined by \citep{zaidiDirectNumericalSimulation2014}:
\begin{equation}
    d_{NN} (i) = \mathop {\min }\limits_{j = 1,2,...,{N_p},j \ne i} \left\| {{{\boldsymbol{x}}_i} - {{\boldsymbol{x}}_j}} \right\|,
\end{equation}
where ${\boldsymbol{x}}_i$ is the centroid position of the $i$-th particle.
In figure \ref{fig:NND_compare}, the p.d.f.s of $d_{NN}$ at different volume fractions are provided, with a comparison with that of the randomly distributed particles.
It is observed that the statistical distribution of $d_{NN}$ shifts to the side of smaller values in all cases. This observation indicates that the dispersed particles become more locally crowded than the randomly distributed particles, which is consistent with the clustering nature of particles manifested by the Voronoi analysis (see figure \ref{fig:Voronoi}).
In addition, the probability of finding touching particles with a center-to-center distance $d_{NN}=2b$ is increased, which can be ascribed to the DKT-like interactions between nearby particles. Especially, there is a noticable secondary peak of the p.d.f. of $d_{NN}$, evaluated at $d_{NN}=2b$, in the most dilute suspension at $\phi=0.1\%$, revealing the prevalence of touching particle pairs undergoing the kissing stage of the DKT process. The touching particle pairs, however, would become less stable with the intensified hydrodynamic disturbances and more frequent inter-particle collisions as $\phi$ increases. As a result, this secondary peak of the p.d.f. of $d_{NN}$ becomes invisible at higher volume fractions. 
Moreover, we also calculate the ensemble average of $d_{NN}$, denoted by $\langle d_{NN} \rangle$, of each case (not presented here). The results show that in the most dilute case at $\phi=0.1\%$, the value of $\langle d_{NN} \rangle$ is $4.0D_{eq}$, considerably larger than that of the case with the strongest particle clustering (i.e. $\langle d_{NN} \rangle=1.9 D_{eq}$ at $\phi=1\%$). 
To make a fair comparison, we normalize $\langle d_{NN} \rangle$ by that of a particle assembly with a random spatial distribution (denoted by $\langle d_{NN} \rangle_{rand}$), and obtain $\langle d_{NN} \rangle/\langle d_{NN} \rangle _{rand}=0.93$ for $\phi=0.1\%$ and 0.70 at $\phi=1\%$. This indicates that particles are closer to the random distribution at the lowest volume fraction, consistent with the attenuation of the particle clustering as $\phi$ decreases from 1\% to 0.1\% obtained by the Voronoi analysis. However, this observation is in a qualitative disagreement with the intensified clustering trend for settling spherical particles at $Ga=178$ with the volume fraction decreasing from $\phi=0.5\%$ to $\phi=0.05\%$ \citep{doychevDynamicsFinitesizeSettling}. We speculate that the disagreement is related to the weaker fluid inertia effect (characterized by the lower value of $Ga$ and also $Re_t$) in the present study. The wake-induced hydrodynamic interactions, which are essential for the attraction among settling particles, may be not strong enough to make sparsely distributed particles to form clusters at very low volume fraction. This argument, however, needs to be further examined by the simulations with the volume fraction lower than 1\%. 

\begin{figure}
    \centering
    \includegraphics[width=0.45\textwidth]{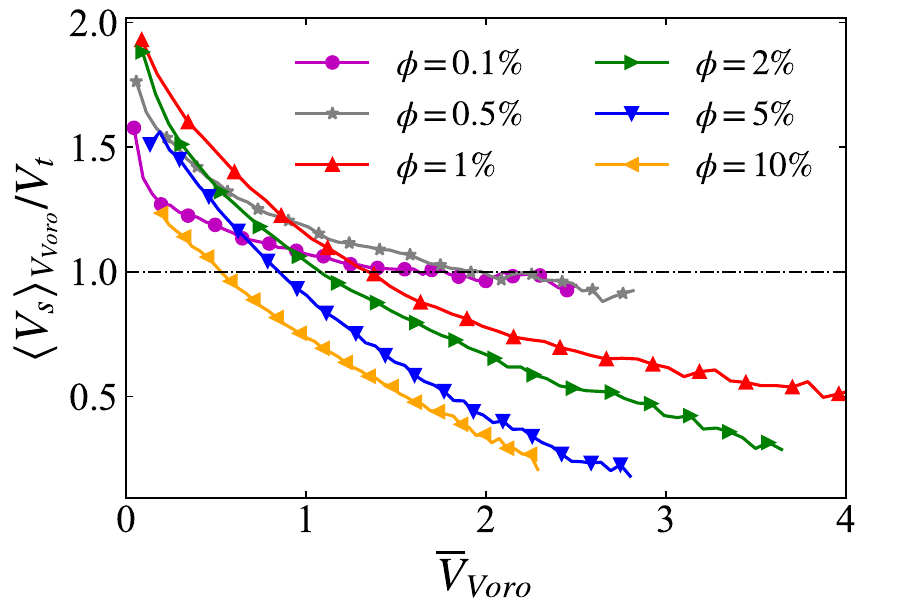}
    \caption{Averaged settling velocity of dispersed particles conditioned on the Voronoi volume at different volume fractions.}
    \label{fig:vsett_voro}
\end{figure}

\begin{figure}
    \centering
    \includegraphics[width=0.99\linewidth]{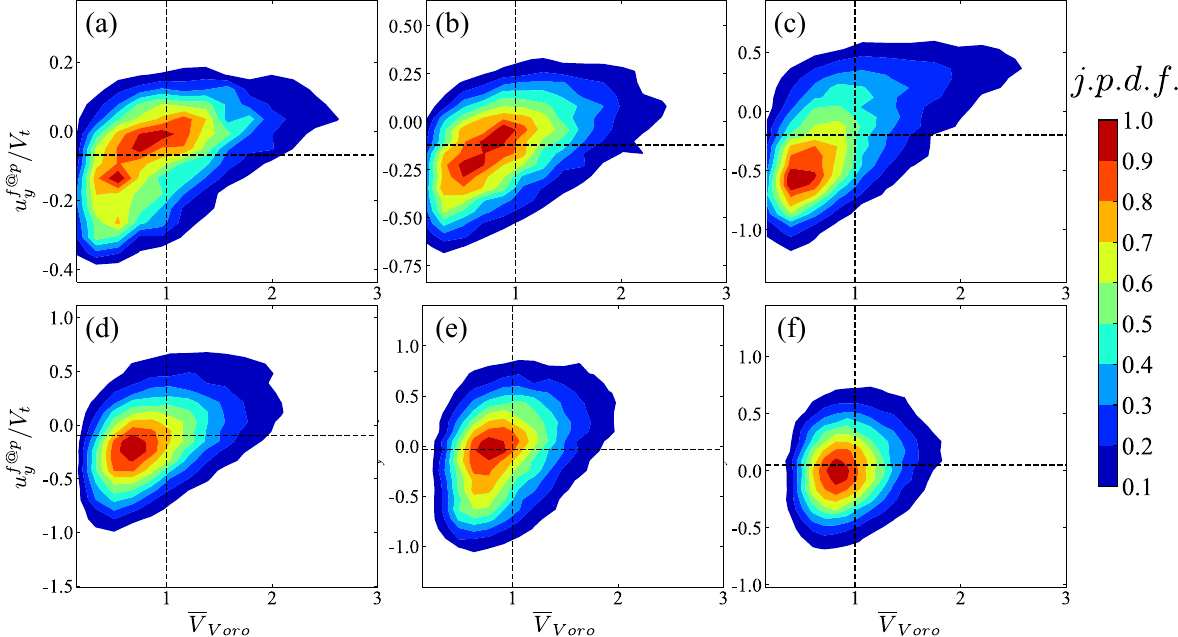}
    \caption{Joint probability density function (j.p.d.f.) (scaled by its maximum value) of the normalized Voronoi volume $\overline{V}_{Voro}$ and the vertical fluid velocity sampled by particles $u_y^{f@p}$ at (a) $\phi=0.1\%$; (b) $\phi=0.5\%$; (c) $\phi=1\%$; (d) $\phi=2\%$; (e) $\phi=5\%$ and (f) $\phi=10\%$. The horizontal and vertical dashed lines represent the mean value of $u_y^{f@p}$ and $\overline{V}_{Voro}$, respectively.}
    \label{fig:jpdf_voro_uy}
\end{figure}
Then, we further investigate the relationship between the clustering and settling velocity of particles in the present flow system. 
\citet{moricheClusteringLowaspectratioOblate2023} reported the positive correlation between the standard deviation of Voronoi volumes and the mean settling velocity of low-aspect-ratio oblate particles. Here, we compute the averaged settling velocity conditioned on the Voronoi volume, denoted by $\langle V_s \rangle_{V_{Voro}}$, and present the results in figure \ref{fig:vsett_voro}. It is shown that particles with smaller Voronoi tessellations tend to settle faster, irrespective of the volume fraction. This correlation can be explained by the so-called ``swarm effect'' \citep{kochInertialEffectsSuspension2001,wangModelingTotalDrag2022}, which suggests that a cluster of settling particles experiences lower total drag compared to the same number of individual particles. 
To gain further understanding, we compute the fluid velocity sampled by the particle, denoted by $\boldsymbol{u}^{f@p}$, by averaging the local fluid velocity on the surface of a sphere centered at the particle centroid with a radius of $1.5 D_{eq}$ \citep{kidanemariamDirectNumericalSimulation2013}. 
Figure \ref{fig:jpdf_voro_uy} shows the joint probability density function of the Voronoi volume and the sampled vertical fluid velocity of the particle. 
In this figure, we observe a positive correlation between $\overline{V}_{Voro}$ and $u_y^{f@p}$, revealing that the particles in the clustering regions (represented by the small value of $\overline{V}_{Voro}$) are prone to sample downward fluid flows, while in the void zones (represented by the large value of $\overline{V}_{Voro}$) particles tend to experience stronger upward flows. This observation can be attributed to the fact that the clustered particles are more likely to reside in the wake of other particles, where the downward flow is dominated. On the contrary, the fluid moves upwards in the void regions so as to decelerate the particle settling motion. However, this correlation becomes less pronounced at $\phi \geq 5\%$, seemingly due to the disruption of particle wakes and the diminished distinction between the ``wake region'' and ``void region'' in dense suspensions.

\begin{figure}
    \centering
    \includegraphics[width=0.99\linewidth]{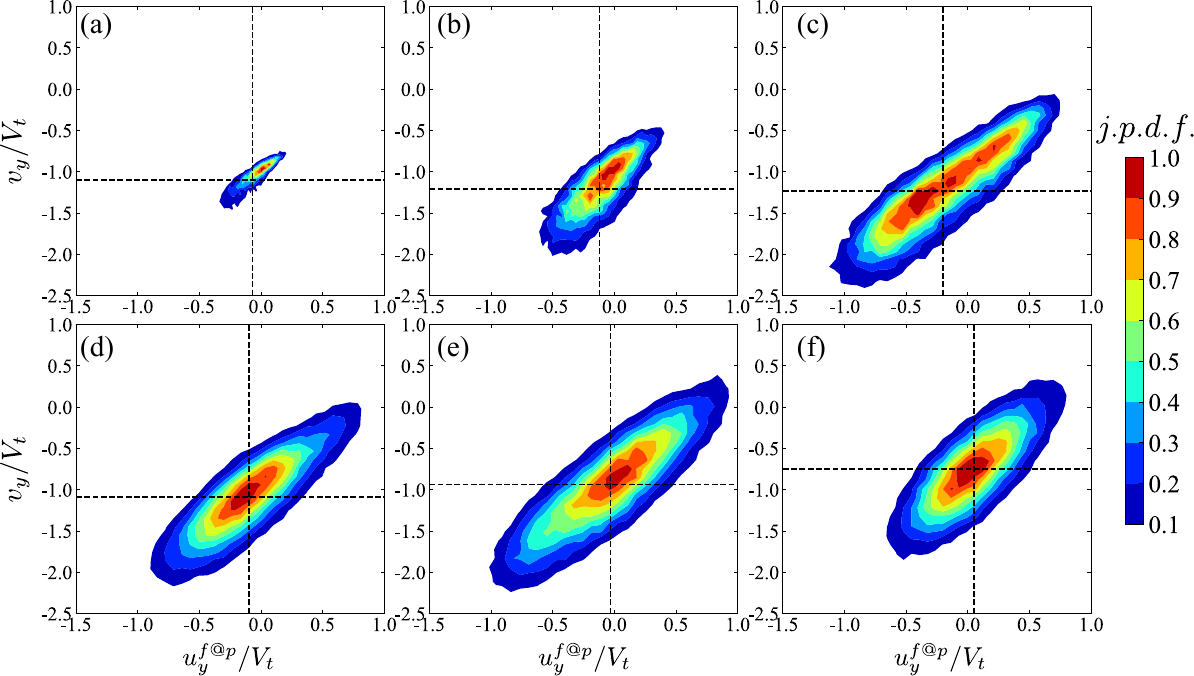}
    \caption{Joint probability density function (scaled by its maximum value) of the vertical fluid velocity sampled by particles, $u_y^{f@p}$, and the particle vertical velocity, $v_y$, at (a) $\phi=0.1\%$; (b) $\phi=0.5\%$; (c) $\phi=1\%$; (d) $\phi=2\%$; (e) $\phi=5\%$ and (f) $\phi=10\%$. The horizontal and vertical dashed lines represent the mean value of $v_y$ and $u_y^{f@p}$, respectively.}
    \label{fig:jpdf_vy_uy}
\end{figure}

\begin{figure}
    \centering
    \includegraphics[width=0.85\linewidth]{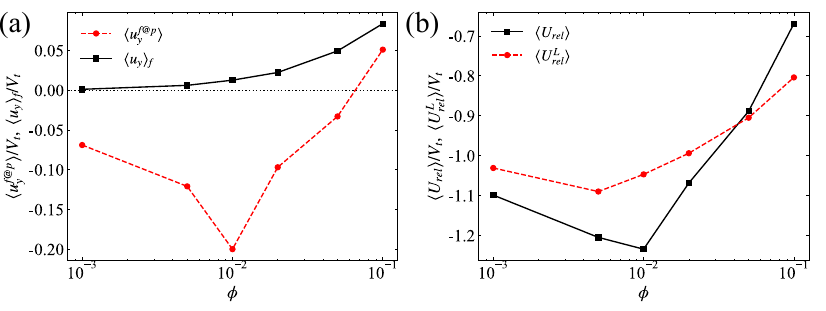}
    \caption{(a) Mean vertical fluid velocity sampled by the particles, $\langle u_y^{f@p} \rangle$, and the ensemble averaged velocity of the fluid flow, $\langle u_y \rangle_f$. 
    (b) Averaged relative velocity between the particle motion and the mean flow, $U_{rel}$, and between the particle motion and the local fluid flow, $U_{rel}^L$.} 
    \label{fig:mean_vel}
\end{figure}
Furthermore, we also examine the relationship between the translational velocity of the particle and the local fluid velocity seen by the particle. In figure \ref{fig:jpdf_vy_uy}, we present the joint probability density function of the vertical component of the particle velocity and the particle-sampled fluid velocity. The results demonstrate that the particles tend to settle rapidly when experiencing vertical downward flows (with the negative value of $\langle u_y^{f@p} \rangle$), and vice versa. The correlations presented in figure \ref{fig:jpdf_voro_uy} and \ref{fig:jpdf_vy_uy} altogether can account for the decreased settling velocity of particles with larger Voronoi volumes shown in figure \ref{fig:vsett_voro}.
In addition, we also compute the average value of the fluid vertical velocity sampled by particles, denoted by $\langle u_y^{f@p} \rangle$, and compare it with the ensemble averaged fluid vertical velocity, $\langle u_y \rangle_f$, in figure \ref{fig:mean_vel}(a). Interestingly, $\langle u_y^{f@p} \rangle$ is always smaller than $\langle u_y \rangle_f$, regardless of the volume fraction. This observation reveals the preferential sampling of downward fluid flows by dispersed particles, which was also reported by \citet{uhlmannSedimentationDiluteSuspension2014} for settling spheres at $Ga=178$. Moreover, the difference between $\langle u_y^{f@p} \rangle$ and $\langle u_y \rangle_f$ is largest at $\phi=1\%$, corresponding to the strongest particle clustering with varying volume fraction. 
Therefore, the preferential sampling of downward flows, which is most significant for the strongest particle clustering, is the underlying mechanism of the aforementioned swarm effect to enhance the particle mean settling velocity.

Additionally, we also look into the relative motion between the particle and fluid phases. Here we define $U_{rel}=v_y-\langle u_y \rangle_f$ as the relative vertical velocity between the particle motion and the mean flow, and $U_{rel}^L=v_y-u_y^{f@p}$ as the local relative velocity. In figure \ref{fig:mean_vel}(b), we provide the ensemble average of these two relative velocities, denoted by $\langle U_{rel} \rangle$ and $\langle U_{rel}^L \rangle$, at different volume fractions. It is observed that the variation of $\langle U_{rel}^L \rangle$ with increasing $\phi$ is alleviated compared to that of $\langle U_{rel} \rangle$. Especially in dilute cases with $\phi \leq 1\%$, the difference of $\langle U_{rel}^L \rangle$ among different cases is less than $4\%$, similar to the observation for settling spheres in dilute suspensions \citep{doychevDynamicsFinitesizeSettling}. Therefore, the variation of the global relative particle-fluid velocity $\langle U_{rel} \rangle$ as $\phi$ changes can be substantially attributed to the different level of particle clustering and the preferential sampling of the fluid velocity. 
More discussion about the variation of $\langle U_{rel}^L \rangle$ is provided in section \ref{subsubsec:orient}. 

\subsubsection{Hindrance effect} \label{subsubsec:hindrance}
Let us now turn to the reduced particle mean settling velocity (i.e. $\langle V_s \rangle < V_t$) in dense suspensions at $\phi \geq 5\%$ (see figure \ref{fig:vel_sett_mean}).
The ensemble averaged fluid velocity, which can be calculated by the flux conservation of the whole system as $\langle u_y \rangle_f=\phi/(1-\phi) \langle V_s \rangle$ \citep{yinHinderedSettlingVelocity2007}, is enhanced as $\phi$ increases (see figure \ref{fig:mean_vel}(a)). The enhanced upward fluid flow has an opposite effect upon the sampling of downward flows by particles. In the meantime, as $\phi$ increases, particle clustering is attenuated (with the clustering indicator $C$ decreasing) and the preferential sampling of downward flows becomes less pronounced (with $\langle u_y^{f@p} \rangle$ approaching $\langle u_y \rangle_f$).
Consequently, the value of $\langle u_y^{f@p} \rangle$ decreases in magnitude when $\phi>1\%$ and even becomes positive at the highest volume fraction $\phi=10\%$. 
As a result, the hindrance effect becomes predominant when the volume fraction exceeds approximately 5\%, leading to the reduction of the mean settling velocity in this regime.

As for the sedimentation of spherical particles, \citet{richardsonSedimentationSuspensionUniform1954} proposed the well-known empirical formula of the hindered settling velocity as a function of the particle volume fraction, i.e.:
\begin{equation} \label{eq:hinder}
    \langle V_s \rangle/V_t=(1-\phi)^n.
\end{equation}
The exponent $n$ in \eqref{eq:hinder} was found to be an decreasing function of the settling Reynolds number $Re_t$ and can be fitted by \citep{garsideVelocityVoidageRelationshipsFluidization1977}:
\begin{equation} \label{eq:exponent}
    \frac{5.1-n}{n-2.7}=0.1Re_t^{0.9}.
\end{equation}
In figure \ref{fig:vel_sett_mean}, we also depict the empirical hindered settling velocity as a function of $\phi$ given by equation \eqref{eq:hinder}, with the exponent $n=3.17$ obtained by substituting $Re_t=61.8$ in equation \eqref{eq:exponent}.
It is shown that the reduced settling velocity observed in the present simulations at $\phi \geq 5\%$ approaches the prediction by the empirical formula \eqref{eq:hinder}. The remaining discrepancy can be ascribed to the weak effect of clustering and the change of orientation (see the discussion on figure \ref{fig:cosy} in the following) of settling prolate particles.

\subsubsection{Particle orientation} \label{subsubsec:orient}
\begin{figure}
    \centering
    \includegraphics[width=0.8\textwidth]{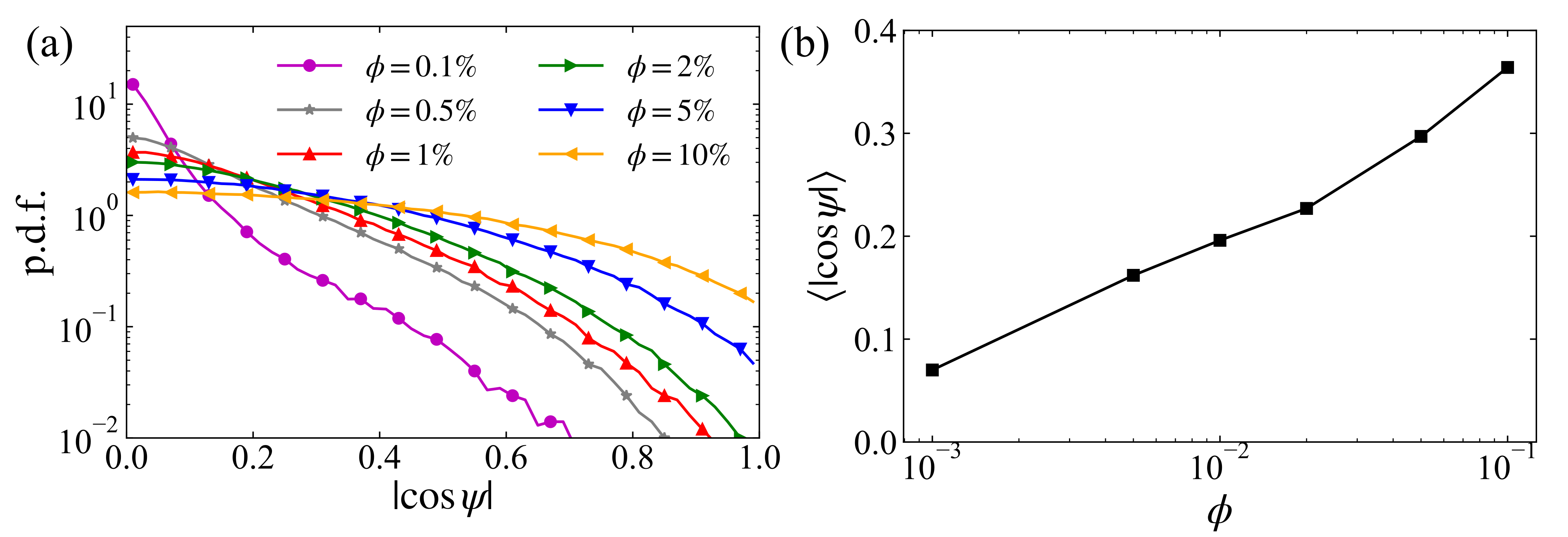}
    \caption{The statistics of the orientation of settling prolate particles at different volume fraction. (a) P.d.f. of the cosine value of the pitch angle $\psi$; (b) Mean value of $|\cos \psi |$ as a function of the volume fraction.}
    \label{fig:cosy}
\end{figure}

\begin{figure}
    \centering
    \includegraphics[width=0.9\linewidth]{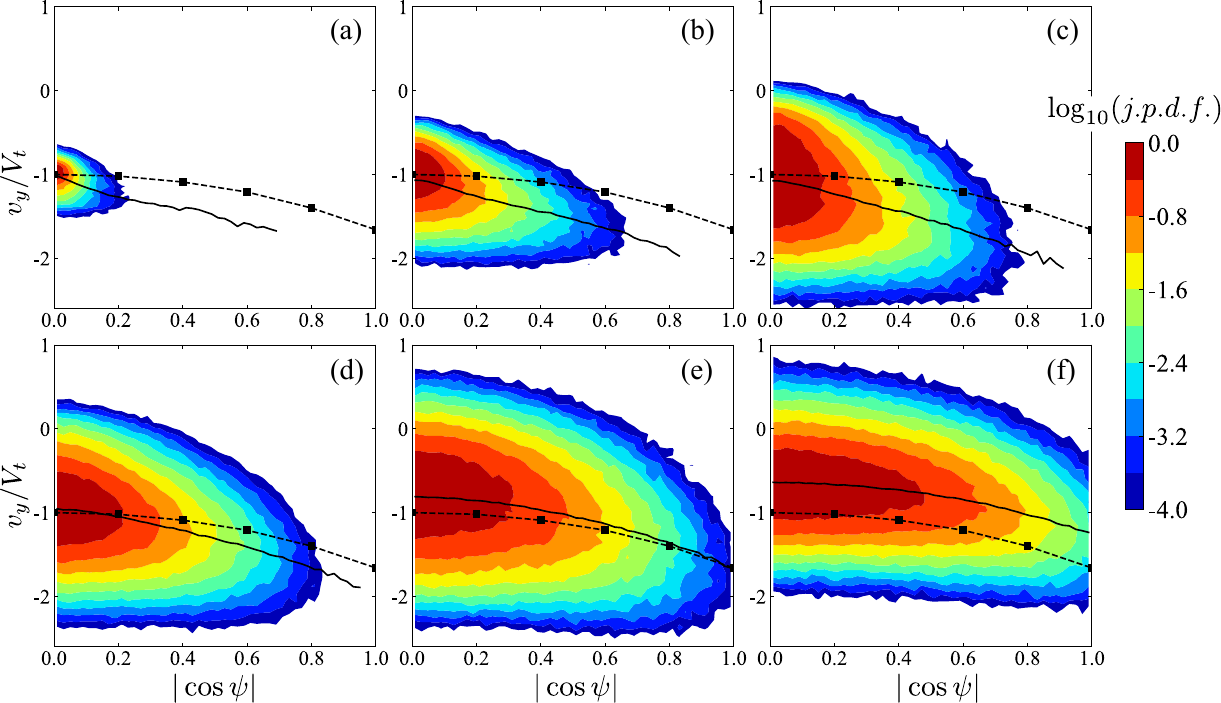}
    \caption{Joint probability density function (scaled by its maximum) of the absolute cosine of the particle pitch angle and the vertical velocity of the particle at (a) $\phi=0.1\%$; (b) $\phi=0.5\%$; (c) $\phi=1\%$; (d) $\phi=2\%$; (e) $\phi=5\%$ and (f) $\phi=10\%$.
    The solid line represents the averaged settling velocity conditioned on the pitch angle (the data with $p.d.f. (| \cos \psi |) <0.01$ are masked). The dashed line represents the variation of an isolated settling prolate spheroid with an artifically fixed pitch angle.}
    \label{fig:jpdf_cos_vy}
\end{figure}

At last, we would like to study the orientation of settling prolate particles and its influence on the particle settling motion.
First, we compute the statistics of particle orientation in the present simulations and display the results in figure \ref{fig:cosy}. It is shown that in the cases with low volume fractions the broad-side-on orientation (corresponding to $|\cos \psi|=0$) of settling prolate spheroids still prevails. This is the stable orientation of an isolated settling prolate spheroid under the effect of the fluid inertia torque \citep{dabadeEffectInertiaOrientation2016,ardekaniNumericalStudySedimentation2016}. However, with the increasing volume fraction, the orientation of particles progressively shifts towards a random distribution, demonstrated by the flattening of the p.d.f. of $|\cos \psi|$ and the corresponding increase in the average value, $\langle|\cos \psi|\rangle$. This observation manifests the overwhelming effect of particle-particle interactions to perturb the stable orientation of settling prolate spheroids in dense suspensions. 
Then, we also examine the correlation between the settling velocity and the orientation of particles in the present flow system by computing the joint probability density function of these two quantities. As shown in figure \ref{fig:jpdf_cos_vy}, prolate particles tend to settle faster as their orientation deviates more from the broad-side-on alignment, irrespective of the volume fraction, just as the case of an isolated settling prolate spheroid. 

\begin{figure}
    \centering
    \includegraphics[width=0.4\linewidth]{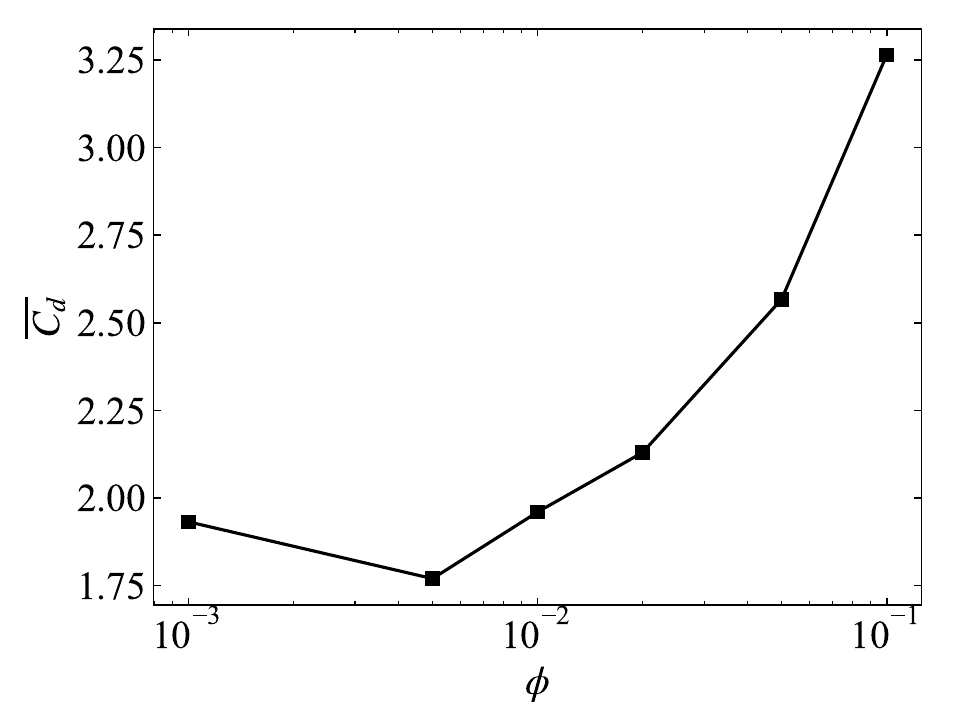}
    \caption{Mean drag coefficient as a function of the particle volume fraction.}
    \label{fig:mean_drag_coeff}
\end{figure}
Furthermore, we examine the influence of particle orientation on the settling motion. 
As for a single spheroid in a uniform flow, the drag coefficient depends on the attack angle between the symmetry axis and the income flow \citep{zastawnyDerivationDragLift2012}. 
To examine the case in the suspension, we compute the mean drag coefficient of the dispersed particles, $\overline {C_d}$, in each case under consideration. The definition of the mean drag coefficient is based on the mean hydrodynamic drag force and the mean local particle-fluid relative velocity, i.e.:
\begin{equation} \label{eq:mean_drag_coeff}
    \overline{C_d}=\frac{\langle F_y^H\rangle}{\frac{1}{2}\rho_f\langle U_{rel}^L\rangle^2A}
\end{equation}
where $F_{y}^H$ is the vertical component of the hydrodynamic force acting on the particle and $A=\pi D_{eq}^2/4$ the characteristic frontal area of the particle.
Note that the time-average of $F_{y}^H$ is in balance with the buoyancy of the particle in the statistically steady state, i.e., $\langle F_y^H\rangle=\pi (\rho_p-\rho_f)g D_{eq}^3/6$. Thus, according to the definition \eqref{eq:mean_drag_coeff}, the variation of $\langle U_{rel}^L\rangle$ shown in figure \ref{fig:mean_vel}(b) is determined by $\overline{C_d}$ for different $\phi$.
As shown in figure \ref{fig:mean_drag_coeff}, $\overline{C_d}$ first decreases slightly as $\phi$ increases from $0.1\%$ to $0.5\%$, which could be attributed to the increased deviation from the broad-side-on orientation of settling prolate particles. However, $\overline{C_d}$ then increases monotonically with the volume fraction when $\phi > 0.5\%$. This cannot be explained by the change of particle orientation since the dispersed particles deviate more from the broad-side-on orientation as $\phi$ continues to grow (see figure \ref{fig:cosy}). 
Therefore, we would like to conclude that the change of particle orientation plays a minor role in the mean settling velocity of dispersed particles, although the settling velocity of individual particles is still orientation-dependent.
Incidentally, the increasing trend of the mean drag coefficient with the increasing particle volume fraction was also reported in the studies of the flow past a fixed or mobile assembly of spherical particles \citep{tennetiDragLawMonodisperse2011,tavanashadParticleresolvedSimulationFreely2021}. 
This trend can be interpreted by the change of the local flow conditions in the vicinity of the particle. Specifically speaking, with the increase of the volume fraction, the fluctuations of the particle and fluid velocities grow with the intensified particle-particle and particle-fluid interactions (see figure \ref{fig:std_and_skw}(a) for more details). As a consequence, the dispersed particles exposed to turbulent local flows would experience increased unsteady and non-linear drags \citep{fornariSedimentationFinitesizeSpheres2016}, which contributes to the drag enhancement of settling particles.

\subsection{Particle microstructures and particle-fluid interactions} \label{subsec:micro}
\begin{figure}
    \centering
    \includegraphics[width=0.8\textwidth]{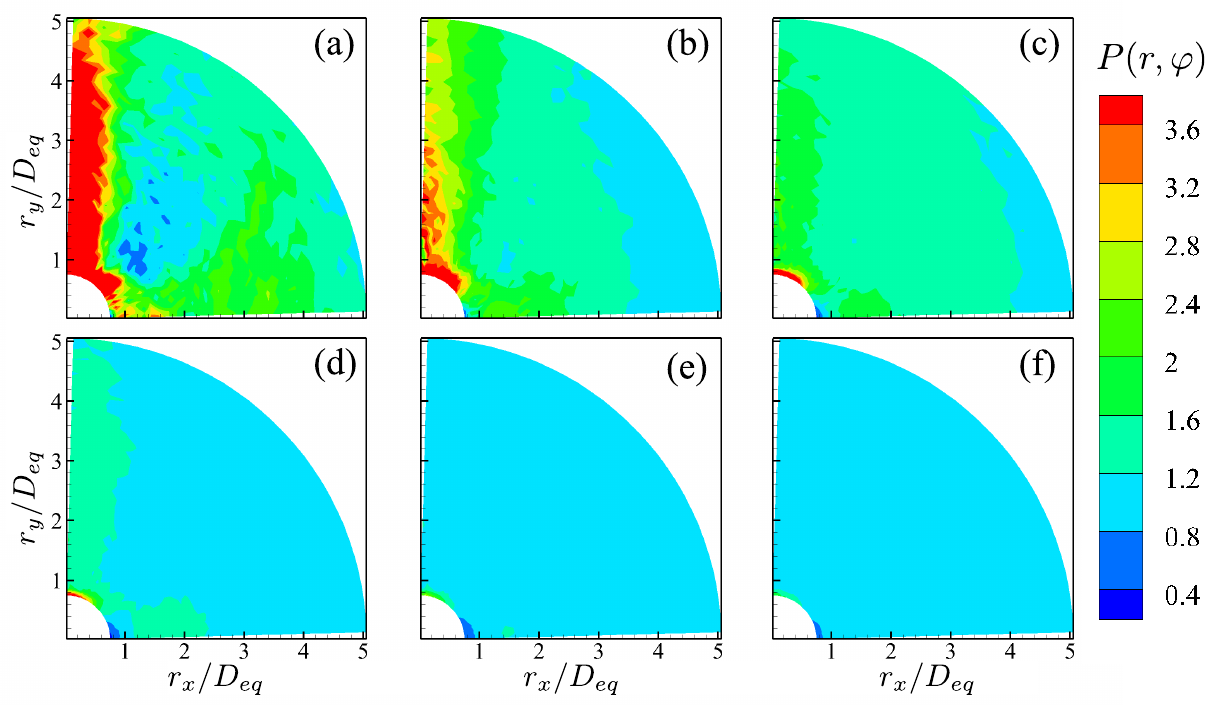}
    \caption{Pair distribution function of particles at (a) $\phi=0.1\%$; (b) $\phi=0.5\%$; (c) $\phi=1\%$; (d) $\phi=2\%$; (e) $\phi=5\%$ and (f) $\phi=10\%$. Here, the horizontal and vertical directions are denoted by $r_x=r \sin \varphi$ and $r_y=r \cos \varphi$, respectively. }
    \label{fig:pairfunc}
\end{figure}

According to the discussion in section \ref{subsubsec:cluster}, the spatial distribution of settling particles is non-uniform in the suspensions. Therefore, it is of interest to examine the particle microstructures. First, we calculate the particle pair distribution function $P(\boldsymbol{r})$, which provides the information about the probability of finding another particle relative to a reference particle with a separation vector $\boldsymbol{r}$. The definition of $P(\boldsymbol{r})$ is referred to \citet{yinHinderedSettlingVelocity2007}, \citet{zaidiDirectNumericalSimulations2015} and \citet{fornariClusteringIncreasedSettling2018}. By definition, $P(\boldsymbol{r}) > 1$ indicates a higher probability of finding a pair of particles with a separation of $\boldsymbol{r}$ compared to the uniform distribution of particles. In addition, as $P(\boldsymbol{r})$ is axisymmetric about the direction of gravity, we calculate the average of $P(\boldsymbol{r})$ over the isotropic horizontal plane (i.e. $x-z$ plane), and obtain the pair distribution function as a function of the separation distance $r=\| \boldsymbol{r} \|$ and the polar angle $\varphi$ (the angle between the positive $y$ direction and the vector $\boldsymbol{r}$), i.e.:
\begin{equation}
    P\left( {r,\varphi } \right) = \frac{1}{{2\pi }}\int_0^{2\pi } {P\left( {\boldsymbol{r}} \right)d\theta }  = \frac{1}{{2\pi }}\int_0^{2\pi } {P\left( {r,\theta ,\varphi } \right)d\theta }.
\end{equation}
Here, the variable of integration $\theta$ is the azimuth angle between the positive $x$ direction and the projection of vector $\boldsymbol{r}$ onto the horizontal plane.

\begin{figure}
    \centering
    \includegraphics[width=0.8\textwidth]{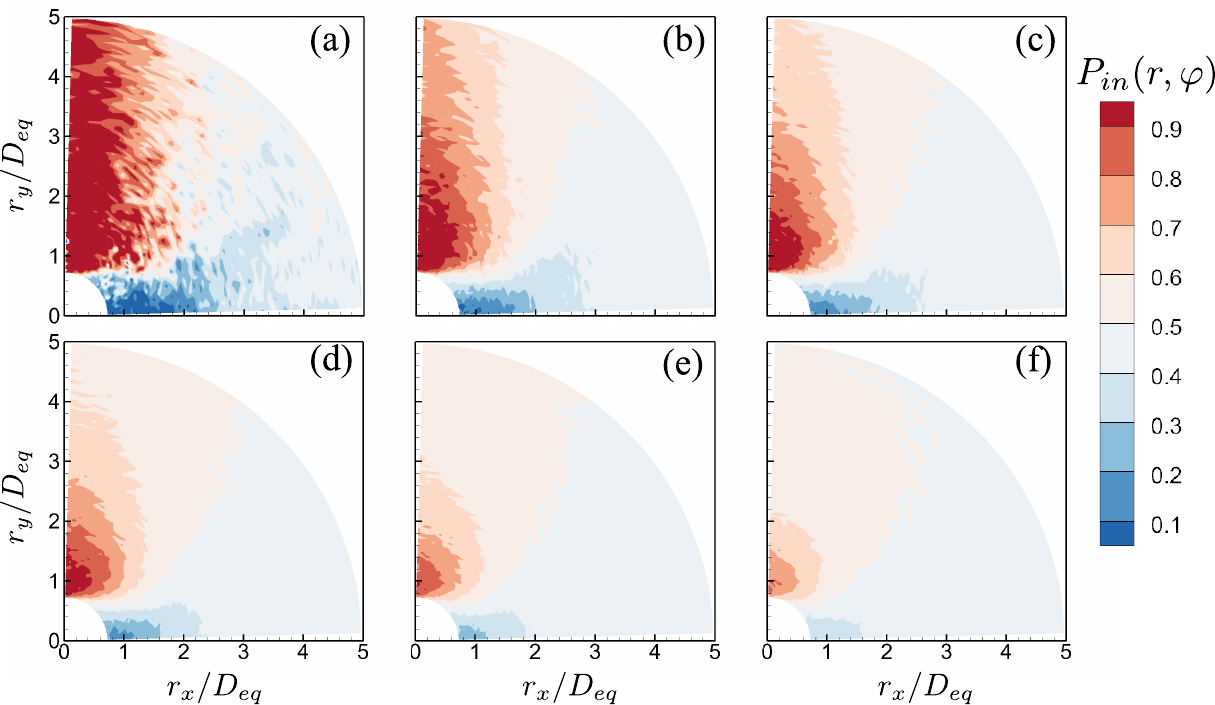}
    \caption{Probability of observing attracting particle pairs as a function of the separation $\boldsymbol{r}$. (a) $\phi=0.1\%$; (b) $\phi=0.5\%$; (c) $\phi=1\%$; (d) $\phi=2\%$; (e) $\phi=5\%$; (f) $\phi=10\%$.} 
    \label{fig:pairin}
\end{figure}

In figure \ref{fig:pairfunc}, we display the particle pair distribution function at different volume fractions. It is observed that the value of $P\left( {r,\varphi } \right)$ is significantly greater than unity along the vertical direction in dilute suspensions. Therefore, we can infer that the dispersed particles prefer to form column-like structures.
To gain more information, we also compute the probability of observing attracting particle pairs with the separation vector $\boldsymbol{r}$, denoted by $P_{in}(\boldsymbol{r})$. Here, an attracting particle pair is identified if the relative radial velocity $W_r$ between the two particles is negative, i.e.:
\begin{equation}
    W_r^{\left\{ {i,j} \right\}} = \frac{{\left( {{{\boldsymbol{v}}_i} - {{\boldsymbol{v}}_j}} \right) \cdot \left( {{{\boldsymbol{x}}_i} - {{\boldsymbol{x}}_j}} \right)}}{{\left\| {{{\boldsymbol{x}}_i} - {{\boldsymbol{x}}_j}} \right\|}}<0,
\end{equation}
where $i$ and $j$ represent the indices of the two particles. Same as the particle pair function, we compute the average of $P_{in}(\boldsymbol{r})$ over the isotropic horizontal directions and obtain the two-dimensional function of $P_{in}(r,\varphi)$, as shown in figure \ref{fig:pairin}. In dilute suspensions, particle pairs exhibit a strong tendency to attract each other along the vertical direction (indicated by $P_{in}>0.5$), but behave in a repulsive manner along the horizontal direction. We attribute these observations to the DKT-like interactions among settling prolate particles. As a result, the attraction and entrapment of particles in the wake of leading ones give rise to the column-like particle microstructures, consistent with the results shown in figure \ref{fig:pairfunc}.
In contrast to our result, it was reported that settling spheres tend to form particle deficits along the vertical direction, while the spatial distribution of cubic particles is more uniform under similar conditions ($Re_t<70$ and $\phi \approx 1\%$) \citep{yinHinderedSettlingVelocity2007,zaidiParticleVelocityDistributions2018,seyed-ahmadiSedimentationInertialMonodisperse2021}. 
These differences highlight the effect of particle shape on the wake-induced hydrodynamic interactions among settling particles. For prolate spheroids, the attraction between particle pairs in the DKT-like events is strong enough to entrap particles in the wake regions \citep{ardekaniNumericalStudySedimentation2016}. Regarding spherical particles, the shear-induced lift force dominates in wake regions, pushing trailing particles outside the wake \citep{yinHinderedSettlingVelocity2007}. While, settling cubic particles are found to have greater rotational rates, which generate lateral Magnus forces that help them escape from the wake of leading particles. Hence, cubic particles are less likely to form obvious microstructures \citep{seyed-ahmadiDynamicsWakesFreely2019,seyed-ahmadiSedimentationInertialMonodisperse2021}. 
Additionally, due to the disruption of particle wakes, the probability of observing attracting/repelling particle pairs is reduced, and the regions where these events dominate diminish with the increasing volume fraction (see figure \ref{fig:pairin}). As a result, the column-like microstructures are gradually attenuated and eventually becomes negligible at $\phi \ge 5\%$.

\begin{figure}
    \centering
    \includegraphics[width=0.8\textwidth]{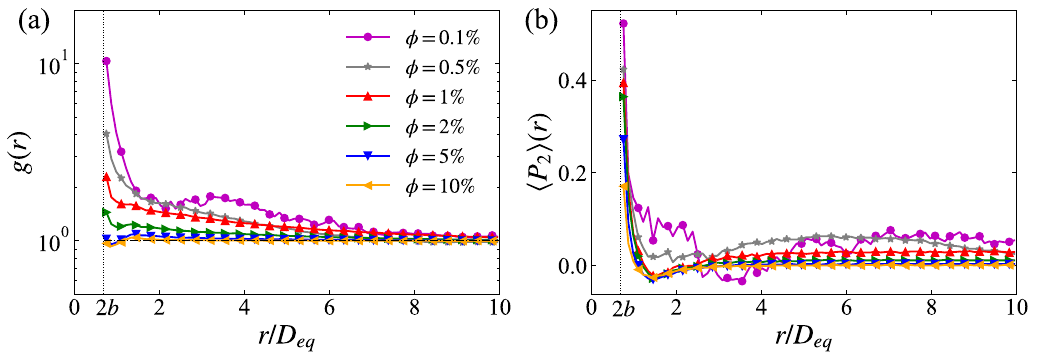}
    \caption{(a) Radial distribution function and (b) order parameter of particle pairs as functions of the radial separation distance $r$. }
    \label{fig:rdf_order}
\end{figure}

Furthermore, we quantify the radial characteristics of particle microstructures by computing the radial distribution function (RDF) of the dispersed particles. 
The RDF, denoted by $g(r)$, is calculated from the two-dimensional pair distribution function $P(r,\varphi)$ by \citep{yinHinderedSettlingVelocity2007}: 
\begin{equation}
    g\left( r \right) = \frac{1}{2}\int_0^\pi  {P\left( {r,\varphi } \right)\sin \varphi d\varphi }.
\end{equation}
As shown in figure \ref{fig:rdf_order} (a), the spatial correlation of particle pairs is considerable increased with $g(r)>1$ for small separation distance $r$ in dilute suspensions.
As the separation distance $r$ grows, the RDF gradually decays to $g(r) \sim 1$, indicating the recovery to the uniform distribution for particle pairs with long-distance separations. 
The peak value of $g(r)$, which is evaluated at the separation distance $r=2b$ at $\phi \le 2\%$, is a monotonically decreasing function of the volume fraction. 
While, in dense suspensions with $\phi \ge 5\%$, the RDF is close to unity for all separation distance, corresponding to the attenuated particle microstructures in these cases.

In previous studies on settling particles, the degree of particle clustering is usually quantified by the maximum value of RDF \citep{zaidiDirectNumericalSimulation2014,fornariClusteringIncreasedSettling2018,zaidiParticleResolvedDirect2018}. However, in the present work, we find that the monotonic decrease of the maximum value of $g(r)$ with the increasing volume fraction does not align with the non-monotonic variation of the clustering indicator based on the Voronoi analysis (see figure \ref{fig:Voronoi} (b)). To interpret this inconsistency, which primarily occurs for $\phi<1\%$, we recall the statistics of the NND provided in figure \ref{fig:NND_compare}. By comparing these two statistics, the RDF and the NND, we attribute the peak of the RDF at $r=2b$ to the prevalence of touching particles pairs in the suspensions, corresponding to the rise of the p.d.f. of $d_{NN}$ at $d_{NN}=2b$. In other words, the RDF actually reflects the pairwise information of the dispersed particles, instead of the particle clustering which is a concept in a global sense. 
The subtle difference between these two quantities becomes especially evident in the most dilute case with $\phi=0.1\%$, where the presence of touching particles involved in individual DKT events (see figure \ref{fig:instant} (g,h)) significantly increases the value of $g(r)$ at $r=2b$ and gives rise to the secondary peak of the p.d.f. of $d_{NN}$ at the same distance. Therefore, we shall argue that the maximum value of the RDF is not a proper criterion to measure particle clustering in the present flow system. 

In addition, we also compute the order parameter $\langle P_2 \rangle (r)$ to measure the orientational feature of the particle microstructures. The order parameter is defined as the angular average of the second Legendre polynomial \citep{yinHinderedSettlingVelocity2007, fornariClusteringIncreasedSettling2018}:
\begin{equation}
    \left\langle {{P_2}} \right\rangle (r) = \frac{{\int_0^\pi  P (r,\varphi ){P_2}(\cos \varphi )\sin \varphi {\rm{d}}\varphi }}{{\int_0^\pi  P (r,\varphi )\sin \varphi {\rm{d}}\varphi }},
\end{equation}
in which ${P_2}\left( {\cos \varphi } \right) = \left( {3{{\cos }^2}\varphi  - 1} \right)/2$. The value of $\langle P_2 \rangle (r)$ would be equal to 1 if all particle pairs with the separation $r$ are vertically aligned, 0 for the isotropic arrangement, and -1/2 if the particle pairs are horizontally aligned \citep{yinHinderedSettlingVelocity2007}. 
In figure \ref{fig:rdf_order} (b), we depict the order parameter as a function of the radial separation distance $r$ at different volume fractions. 
In all cases under consideration, the order parameter is greater than zero at $r=2b$, indicating the tendency of nearby particle pairs to align vertically. This phenomenon again manifests the DKT-like hydrodynamic interactions among settling particles in the present system, which also make difference even at $\phi\geq5\%$ with weak particle clustering.
While, the peak value of $\langle P_2 \rangle$ decreases as the volume fraction increases, indicating the weakening influence of particle wakes. Additionally, $\langle P_2 \rangle (r)$ decays with the separation distance $r$ rapidly to $\langle P_2 \rangle \sim 0$, indicating the recovery to an isotropic particle arrangement, in the suspensions with $\phi \ge 2\%$. 
However, for the lower volume fractions, the order parameter does not completely decay to zero with a small residual positive value even for the long-distance particle separation of $r\sim 10 D_{eq}$. This indicates that the wake-induced hydrodynamic interactions have a long working distance for sparsely distributed particles in dilute suspensions.

\begin{figure}
    \centering
    \includegraphics[width=0.99\textwidth]{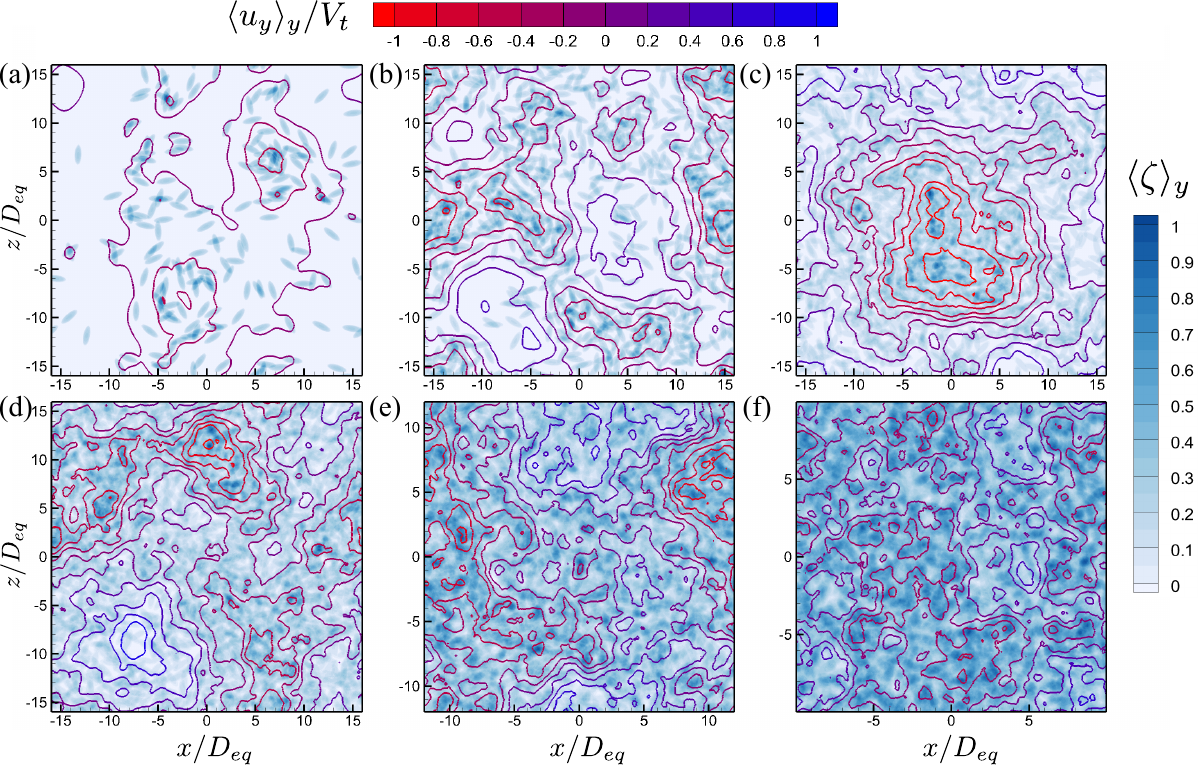}
    \caption{Instantaneous particle concentration $\langle \zeta \rangle_y$ (scaled by its maximum and represented by the background colored contour) and fluid vertical velocity $\langle {{u_y}} \rangle_y$ (represented by the contour lines) averaged over the vertical direction at different volume fraction. (a) $\phi=0.1\%$; (b) $\phi=0.5\%$; (c) $\phi=1\%$; (d) $\phi=2\%$; (e) $\phi=5\%$; (f) $\phi=10\%$.} 
    \label{fig:2D_horizontal}
\end{figure}

Moreover, to demonstrate the fluid-particle interactions in the present flow system, we examine the instantaneous vertical average of the particle concentration (denoted by $\langle \zeta \rangle_y$) and fluid vertical velocity (denoted by $\langle {{u_y}} \rangle_y$) at different volume fractions. The definitions of $\langle \zeta \rangle_y$ and $\langle {{u_y}} \rangle_y$ are as follows:
\begin{align}
    \langle \zeta \rangle_y  \left( {x,z} \right) &= \frac{1}{{{L_y}}}\int_0^{{L_y}} {\zeta \left( {x,y,z} \right)dy}, \\
    \langle {{u_y}} \rangle_y \left( {x,z} \right) &= \frac{1}{{{L_y}}}\int_0^{{L_y}} {\left[ {1 - \zeta \left( {x,y,z} \right)} \right]{u_y}\left( {x,y,z} \right)dy}.
\end{align}
Here, $\zeta(x,y,z)$ is an indicator function which is equal to 1 if a spatial point $(x,y,z)$ locates inside a particle, otherwise $\zeta(x,y,z)=0$.
As shown in figure \ref{fig:2D_horizontal}, the distribution of $\langle \zeta \rangle_y$ on the horizontal plane is far from uniform in dilute suspensions, indicating the formation of column-like particle microstructures. Also, we can evidently observe spatial correlation between the high value of $\langle \zeta \rangle_y$ and the extreme negative value of $\langle {{u_y}} \rangle_y$, and vice versa. In the regions where particles accumulate, the fluid flow moves downwards due to the drag by settling particles, but moves upwards in the regions devoid of particles to keep zero net flux of the whole system. However, as the volume fraction increases, the structures for $\langle \zeta \rangle_y$ and $\langle {{u_y}} \rangle_y$ become fragmented and the above-mentioned correlation is weakened, owing to the diminishing particle microstructures in dense suspensions.

\begin{figure}
    \centering
    \includegraphics[width=0.8\linewidth]{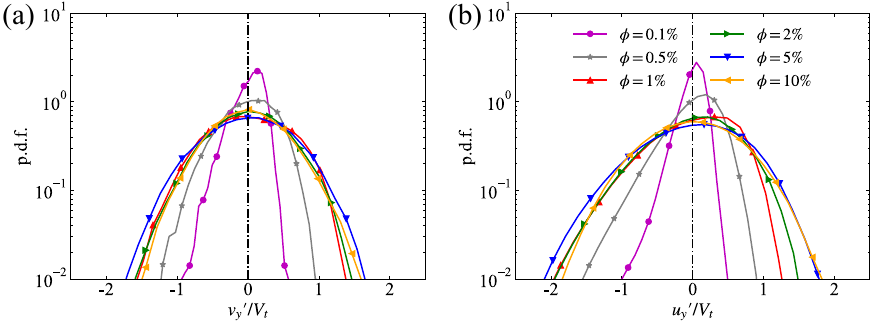}
    \caption{P.d.f. of the vertical component of (a) the particle velocity fluctuations, $v_y'=v_y-\langle v_y \rangle$, and (b) the fluid velocity fluctuations, $u_y'=u_y-\langle u_y \rangle_f$.} 
    \label{fig:pdf_vy_uy}
\end{figure}

\begin{figure}
    \centering
    \includegraphics[width=0.8\linewidth]{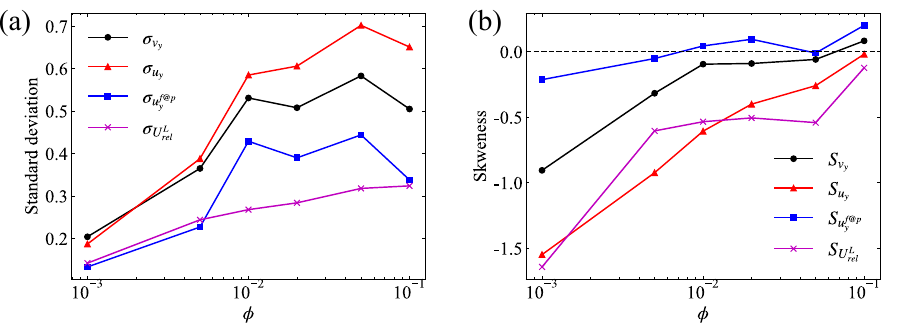}
    \caption{(a) Standard deviation (normalized by $V_t$) and (b) skewness of the vertical component of the particle velocity $v_y$, the fluid velocity $u_y$, the fluid velocity sampled by the particle $u^{f@p}_y$ and the relative velocity between the particle motion and the local fluid flow $U^L_{rel}$ }
    \label{fig:std_and_skw}
\end{figure}

At last, we look into the statistics of the velocity fluctuations of the particle and fluid phases under the effect of the complicated particle-fluid interactions. In figure \ref{fig:pdf_vy_uy}, we illustrate the probability density functions of the fluctuation velocity of the particle and fluid vertical motions. 
In general, the magnitudes of the velocity fluctuation of the two phases, which are quantified by the standard deviations $\sigma_{v_y}$ and $\sigma_{u_y}$ (depicted in figure \ref{fig:std_and_skw}(a)), increase with the volume fraction. 
Since the velocity fluctuation of the particle motion can be decomposed to the contributions from the particle-sampled fluid flow and the local particle-fluid relative motion, i.e. $v_y'={U^L_{rel}}'+{u_y^{f@p}}'$, we have $\sigma_{v_y}^2 \approx \sigma_{u^L_{rel}}^2+ \sigma_{u_y^{f@p}}^2$ (with a negligible residual due to the crossing term of these two contributions) \citep{uhlmannSedimentationDiluteSuspension2014}.
Therefore, to gain further understanding, we also present the information about the second and third moments of $U^L_{rel}$ and $u_y^{f@p}$ in figure \ref{fig:std_and_skw}. 
As shown in figure \ref{fig:std_and_skw}(a), the contributions from $\sigma_{u_y^{f@p}}$ and $\sigma_{u^L_{rel}}$ to the particle velocity fluctuation are almost equal at $\phi=0.1\%$, 0.5\% and 10\%. However, interestingly, the former contribution dominates the latter one in the other three cases, and this tendency is most significant at $\phi=1\%$, corresponding to the strongest particle clustering. This observation was also discussed in \citet{uhlmannSedimentationDiluteSuspension2014} and can be attributed to the different local fluid velocity experienced by the particles in the clustering and void regions. Regarding the the local particle-fluid relative motion, we find that $\sigma_{u^L_{rel}}$ increases monotonically as $\phi$ increases. This increase of fluctuation is presumably due to the influence of the greater turbulence intensity (manifested by the increase of $\sigma_{u^y}$) on the particle-fluid relative motion in dense suspensions.

In addition, the asymmetric distribution of the particle and fluid velocity fluctuations shown in figure \ref{fig:pdf_vy_uy} is worthy of further discussion. First, the asymmetry of the p.d.f. of both
$v_y'$ and $u_y'$, manifested by the higher negative tails, is appreciable in dilute cases. 
The skewed distribution of the fluid vertical velocity fluctuation, which has been reported in the flow induced by settling particles \citep{doychevDynamicsFinitesizeSettling} or rising bubbles \citep{rissoAgitationMixingTransfers2018}, is related to dominant effect of wakes. 
With the increase of the volume fraction, the p.d.f.s of velocity fluctuations gradually recover to be symmetric, which is confirmed by the decrease of the skewness in magnitude shown in figure \ref{fig:std_and_skw}(b) and indicates the attenuated effect of the particle wakes. 
Second, we notice from figure \ref{fig:std_and_skw}(b) that the skewness of the particle-sampled fluid velocity, $S_{u^{f@p}_y}$, is negligible compared to that of the local fluid-particle relative velocity, $S_{U^L_{rel}}$. Therefore, the asymmetric distribution of the particle vertical velocity at low volume fractions seems to be mainly related to the local particle-fluid motion. We speculate that the considerable variation of $S_{U^L_{rel}}$ with varying volume fraction may be caused by the change of the flow condition in the vicinity of the dispersed particles, which is worthy of further investigation.

\subsection{Particle-particle collisions} \label{subsec:collision}

We finally investigate the collision rate of settling particles in the present flow system. Here, we introduce the collision kernel $\Gamma$ to quantify the collision rate. The dynamic collision kernel, denoted by $\Gamma^D$, is defined by \citep{wangStatisticalMechanicalDescription2000,wangTheoreticalFormulationCollision2005}:
\begin{equation}
    \Gamma^D = \frac{2\dot{N}_C}{n^2},
\end{equation}
where $\dot{N}_C$ represents the number of collision events per unit volume per unit time, and $n=N_p/V_{tot}$ is the number density of particles in the suspension. 
In the meantime, the collision kernel can also be described in a kinematic form (namely the kinematic collision kernel $\Gamma^K$), i.e. the inward flux of particles across the surface of a sphere with a collision radius $R_{12}$, as \citep{wangStatisticalMechanicalDescription2000}:
\begin{equation} \label{eq:gamma_k}
    {\Gamma ^K} = 2\pi R_{12}^2 \langle |W_r(R_{12})| \rangle  g\left( {{R_{12}}} \right).
\end{equation}
We notice that the particle pair statistics are involved in the above definition. Specifically, $\langle |W_r(R_{12})| \rangle$ represents the average absolute radial relative velocity (RRV) of particle pairs with a center-to-center distance $R_{12}$, and $g(R_{12})$ is the RDF evaluated at the collision radius $R_{12}$.
It is important to note that the definition \eqref{eq:gamma_k} is valid only under the flux-balance assumption \citep{wangStatisticalMechanicalDescription2000}, which requires the equality between the inward and outward fluxes of the particle motion across the surface of the collision sphere.
For clarity, we define the averaged magnitude of the negative and positive radial relative velocity as the inward and outward velocity $\langle W_r^ -\rangle$ and $\langle W_r^ +\rangle$, respectively. Then, the average absolute RRV can be expressed as:
\begin{equation}
    \langle |W_r| \rangle=P_{in}\langle W_r^ -\rangle +(1-P_{in})\langle W_r^ +\rangle,
\end{equation}
where $P_{in}$ is the probability of observing negative the radial relative velocity (as has been shown in figure \ref{fig:pairin}). 
Accordingly, the flux-balance assumption is valid when the ratio between the inward and outward flux, $C_p$, defined by \citep{wangStatisticalMechanicalDescription2000}:
\begin{equation}
    {C_p} = \frac{{{P_{in}}\langle W_r^ -\rangle }}{{\left( {1 - {P_{in}}} \right)\langle W_r^ +\rangle }},
\end{equation}
is equal to unity at the separation distance $r=R_{12}$. 

\begin{figure}
    \centering
    \includegraphics[width=0.99\textwidth]{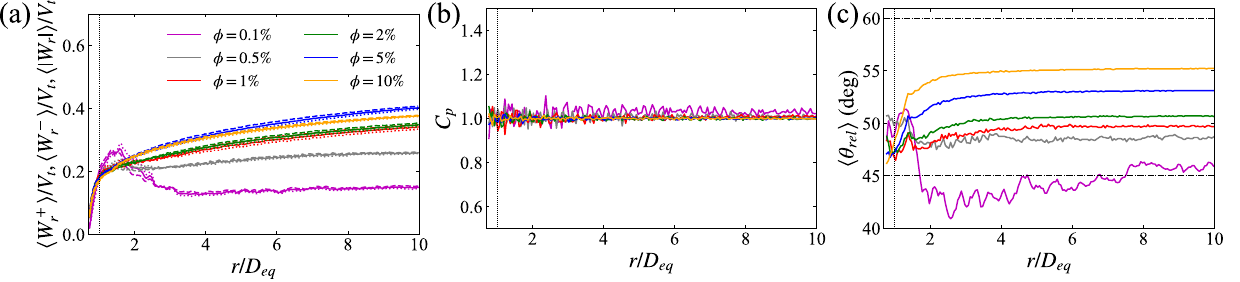}
    \caption{Particle pair statistics as functions of the separation distance $r$. Each line starts from $r=2b$ because of the geometric constraint for finite-size particles. The vertical black dotted line in each panel represents the separation distance $r=D_{eq}$. (a) Relative radial velocity. Dashed lines: average inward velocity $\langle W_r^- \rangle$; dotted lines: average outward velocity $\langle W_r^- \rangle$; solid lines: average absolute velocity $\langle |W_r| \rangle$. (b) The ratio between the inward and outward flux. (c) Average relative angle between particle pairs. The horizontal dash-dotted lines represent $\langle \theta_{rel} \rangle =45^\circ$ and $\langle \theta_{rel} \rangle =60^\circ$.}
    \label{fig:pairStatistics}
\end{figure}

In figure \ref{fig:pairStatistics}, we present the particle pair statistics as functions of the radial separation distance $r$.
To begin with, figure \ref{fig:pairStatistics}(a) displays the relative radial velocities at different volume fractions, leading to the following key observations. First, for a large separation distance $r$, the relative radial velocities generally increase with the particle volume fraction $\phi$, which is ascribed to the intensified particle-fluid and particle-particle interactions as $\phi$ grows. However, the slight decrease of $\langle |W_r| \rangle$ from $\phi=5\%$ to $\phi=10\%$ may be related to the weakening effect of particle wakes, reminiscent of the decrease of the particle and fluid velocity fluctuations shown in figure \ref{fig:std_and_skw}(a). 
Second, a notable peak value of $\langle |W_r| \rangle$ is observed at the separation distance $r\approx 1.5D_{eq}$ in the most dilute case of $\phi=0.1\%$. This reflects the influence of wake-induced DKT-like interactions on the particle relative motion. However, this peak value diminishes as $\phi$ increases due to the disruption of particle wakes. Third, as the separation distance approaches $r=2b$, the relative radial velocities of particles decrease rapidly. The deceleration of the approaching/separating motions between nearby particles should be attributed to the lubrication effect. Last, the mean inward velocity $\langle W_r^ -\rangle$ and outward velocity $\langle W_r^ +\rangle$ exhibit slight discrepancies, indicating that the particle velocity field is compressible \citep{wangStatisticalMechanicalDescription2000}. While, as shown in figure \ref{fig:pairStatistics} (b), the ratio $C_p$ between the inward and outward fluxes is equal to unity (with some fluctuations due to the statistical error), irrespective of the volume fraction. This indicates that the flux-balance assumption remains to be valid in the present four-way coupling particle-fluid system, same as in the one-way coupling simulations of point-particle-laden turbulent flows \citep{wangStatisticalMechanicalDescription2000}.

Moreover, we also illustrate the average relative angle $\langle \theta_{rel} \rangle$ between the symmetry axes of particle pairs with the varying separation distance in figure \ref{fig:pairStatistics} (c). 
Interestingly, $\langle \theta_{rel} \rangle$ seems to converge to approximately $ \langle \theta_{rel} \rangle \approx 45^\circ - 50^\circ$ with the approaching of the particle-particle distance, irrespective of the volume fraction. 
This observation is quite different from the alignment of passive directors in the turbulence \citep{zhaoPassiveDirectorsTurbulence2019}, and may be related to the hydrodynamic interactions between particle pairs \citep{ardekaniNumericalStudySedimentation2016}.
Additionally, as the separation distance increases, the value of $\langle \theta_{rel} \rangle$ saturates to a certain value dependent on the particle volume fraction: The saturated value of $\langle \theta_{rel} \rangle$ is an increasing function of $\phi$, which can be explained by the orientational behavior of settling particles. 
In the limiting case where all particles settle with the major axis perpendicular the gravity, the random orientation of particles on the two-dimensional horizontal plane will lead to $\langle \theta_{rel} \rangle=45^\circ$. The case with $\phi=0.1\%$ is close to this limit as the broad-side-on orientation dominates therein. While, in another limit where the particle orientations are totally random in the three-dimensional space, the average relative angle should be $\langle \theta_{rel} \rangle=60 ^\circ$. This interprets the increase of $\langle \theta_{rel} \rangle$ as $\phi$ increases, in consideration of the randomization of the particle orientation in the dense suspension (see figure \ref{fig:cosy} (a)). 

\begin{figure}
    \centering
    \includegraphics[width=0.8\textwidth]{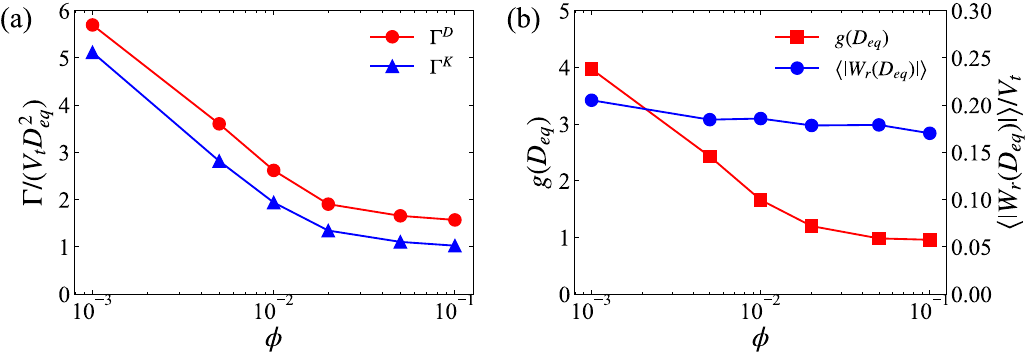}
    \caption{(a) Dynamic and kinematic collision kernels and (b) the RDF and RRV evaluated at $r=D_{eq}$ at different particle volume fraction.}
    \label{fig:collision}
\end{figure}

Now we return to the discussion on the particle collision kernel. When the flux-balance assumption is valid, $\Gamma ^K$ should be strictly equivalent to $\Gamma ^D$ for spherical particles with the collision radius being the diameter of the sphere \citep{wangStatisticalMechanicalDescription2000}.
However, for spheroidal particles, deriving the exact expression of the kinematic collision kernel $\Gamma^K$ is theoretically challenging due to the complexity of their geometry. Alternatively, \citet{siewertCollisionRatesSmall2014} proposed treating the expression (\ref{eq:gamma_k}) as an approximate kinematic collision kernel for spheroidal particles by using the equivalent diameter of the spheroid as the collision radius (i.e. $R_{12}=D_{eq}$).

In figure \ref{fig:collision} (a), we illustrate the dynamic and approximate kinematic collision kernel of settling prolate particles at different volume fractions. The results demonstrate that $\Gamma^K$ underestimates the exact dynamic collision kernel $\Gamma^D$ in the present flow system, similar to the case of settling spheroids in a quiescent fluid with the negligence of particle-particle interactions \citep{jiangFlowreconstructionBasedApproach2024}. Since the flux-balance assumption has been validated, the discrepancy between $\Gamma^K$ and $\Gamma^D$ should be ascribed to the above-mentioned approximation regarding the geometry of the prolate spheroid when defining the approximate kinematic collision kernel. Even though, $\Gamma^K$ still provides a reasonable estimation of the collision kernel, as it qualitatively captures the decreasing trend of $\Gamma^D$ with the increasing volume fraction $\phi$. 

Furthermore, we depict the variation of $g(D_{eq})$ and $\langle | W_r (D_{eq}) | \rangle$, both of which contribute to the kinematic collision kernel $\Gamma^K$ via equation \eqref{eq:gamma_k}, with the change of the volume fraction in figure \ref{fig:collision} (b). 
On the one hand, $g(D_{eq})$ decreases monotonically as $\phi$ increases, which plays an essential role in the reduction of $\Gamma^K$.
This highlights the significance of the wake-induced particle microstructures (as has been discussed in section \ref{subsec:micro}) to the collision rate for settling particles. 
Accordingly, we remark that although the maximum value of RDF is not an appropriate criterion to quantify the particle clustering, the RDF is particularly relevant to the collision efficiency of particles as it directly contributes to the kinematic collision kernel. 
On the other hand, the average absolute RRV, $\langle | W_r (D_{eq}) | \rangle$, exhibits a minor degree of variation with the change of the volume fraction. Previous studies on settling spheroidal particles in turbulence, which ignored particle-particle interactions, reported the enhancement in the average RRV owing to the dispersion of the settling velocity of randomly-oriented spheroidal particles \citep{siewertCollisionRatesSmall2014, juchaSettlingCollisionSmall2018}. However, this mechanism does not apply to the present flow system, although particle orientations become more randomized as $\phi$ increases. 
With the fully-resolved particle-particle hydrodynamic interactions herein, we attribute the abovementioned difference to the predominate effect of lubrication on the relative motion among nearby particles. 
Under the lubrication effect, the relative radial velocity of particle pairs reduces with the separation distance approaching $r=D_{eq}$ (see figure \ref{fig:pairStatistics} (a)), and is responsible for the nearly constant value of $\langle | W_r (D_{eq}) | \rangle$ for different volume fractions.

\section {Concluding remarks} \label{sec:conclud}

\begin{figure}
    \centering
    \includegraphics[width=0.99\textwidth]{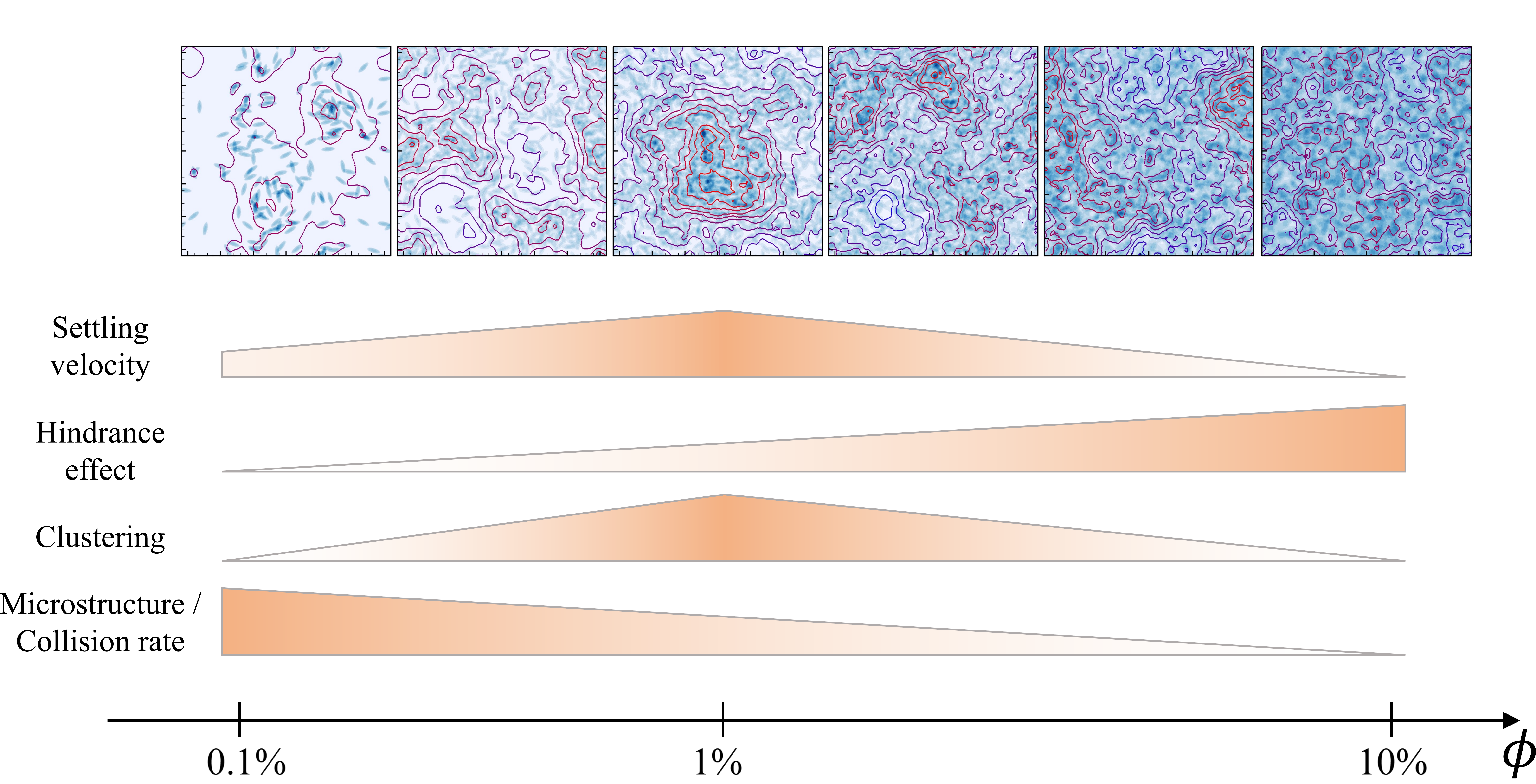}
    \caption{Summary sketch of the findings in the present work regarding settling prolate particles with the varying volume fraction.}
    \label{fig:sketch_summary}
\end{figure}

In the present work, we investigate the sedimentation of prolate particles in a quiescent fluid. Focusing on the volume fraction effect, we conducted the PR-DNS of settling particles in the suspensions with different volume fractions from $\phi=0.1\%$ to $\phi=10\%$. The main findings in the present work are sketched in figure \ref{fig:sketch_summary}. Strikingly, we observe a non-monotonic variation of the mean settling velocity of the dispersed particles with the increase of volume fraction. The highest mean settling velocity is present at an intermediate volume fraction $\phi=1\%$, accompanied with the most significant particle clustering. 
By further investigating the fluid velocity sampled by dispersed particles, we illustrate the preferential sampling of the downward fluid flows for particles in the clustering regions, which underlies the so-called swarm effect and accelerates the settling motion of the dispersed particles. In the cases $\phi<1\%$, the degree of clustering is lowered for sparsely distributed particles, presumably due to the limited effect of wake-induced hydrodynamic attractions among settling particles. 
In another limit with a high volume fraction, however, the crowded arrangement of particles disrupts particle wakes, which also inhibits the formation of particle clustering. In this regime, the enhanced upward mean flow makes the hindrance effect become predominant, and results in the reduction of the particle mean settling velocity to less than the isolated settling velocity. 
In contrast to the particle clustering and hindrance effect, the change of particle orientation plays a minor role in determining the mean settling velocity, although individual prolate spheroids in the suspension still tend to settle faster when then deviate more from the broad-side-on orientation.

In the second part of this study, we investigate the microstructure of settling prolate particles. The study on the particle pair distribution function reveals that particles tend to form vertically aligned microstructures in dilute suspensions. By further looking into the relative velocity of particle pairs with different separation positions, we attribute the presence of column-like particle microstructures to the wake-induced hydrodynamic attractions among settling particles. 
It is noted that although the particle clustering is weakened in the most dilute case with $\phi=0.1\%$, the spatial distribution of particles is far from uniform therein. This is ascribed to the long working distance of the wake-induced hydrodynamic interactions for the sparsely distributed particles in the space.
Under this effect, individual DKT events become prevalent in the dilute suspension, which increases the probability of finding vertically aligned particle pairs, and also augments the radial distribution function at the close separation distance. Additionally, the velocity fluctuations for both the particle and fluid phases exhibit an asymmetric distribution in dilute suspensions due to the complicated particle-fluid interactions.
However, with the increasing particle volume fraction, the microstructures progressively diminish owing to the disruption of particle wakes, and the dispersed particles tends to become uniformly distributed in the space. Meanwhile, the statistical distributions of the fluid and particle velocities recover to be symmetric with an enhanced fluctuations in dense suspensions.

The final part of this study focuses on the collision rate of settling particles. To quantify the collision efficiency, we introduce the collision kernel and discuss this quantity in both dynamic and kinematic perspectives. As the particle pair statistics are involved in the kinematic representation of the collision kernel, we also examine the relative velocity and orientation between particle pairs. 
It is demonstrated that different mechanisms dominate the relative radial velocity of particle pairs at different separation distances: the lubrication effect at short distances, the wake-induced DKT-like interactions at intermediate distances, and the overall fluid-particle and particle-particle interactions at longer distances. 
Despite the compressibility of the particle velocity field, which is indicated by the non-equal inward and outward average velocities for particle pairs, the flux-balance assumption remains to be valid in the present four-way coupling particle-fluid system. 
The angle between particle pairs exhibits similarity for close particle separations, but becomes volume-fraction-dependent at long separation distances due to the different orientational distribution of the dispersed particles. 
As regards the collision efficiency, the monotonic decrease of the collision kernel with the increasing volume fraction is primarily caused by the decrease of the RDF, which is related to the diminishing particle microstructures. This finding highlights the significance of the radial distribution function in determining the particle collision rate, although it cannot provide a reliable measure of the particle clustering. In contrast, the radial relative velocity between particle pairs with the separation distance equal to the collision radius remains almost constant at different volume fractions. 
This is attributed to the predominant lubrication effect among nearby particle pairs, which decelerates their approaching/separating motions at the close distance. Hence, the effect of particle relative motion contributes little to the variation of the collision kernel when the volume fraction changes. In summary, the particle-particle hydrodynamic interactions, including the wake-induced attractions and the lubrication effect, are crucial in affecting the collision rate of settling prolate spheroids.

\backsection[Funding]{This work is supported by the Natural Science Foundation of China (Grant Nos.12388101, 92252104 and 9225220).}

\backsection[Declaration of Interests]{The authors report no conflict of interest.}

\appendix

\section{Numerical method and validations} \label{app:method}
\subsection{Fluid flow simulation and the immersed boundary method}
In the present work, we use the PR-DNS to solve the fluid flow laden with freely-moving particles. Specifically, we adopt the immersed boundary method (IBM) to resolve the particle-fluid interactions \citep{peskinImmersedBoundaryMethod2002,iaccarinoImmersedBoundaryMethod2004}.
We consider the Newtonian fluid with density $\rho_f$ and dynamic viscosity $\mu_f$, and the fluid flow is governed by the incompressible Navier-Stokes (N-S) equations as:
\begin{align}
    \nabla \cdot \boldsymbol{u} &=0 \label{eq:NS_conti}, \\
    \rho_{f}\left(\frac{\partial \boldsymbol{u}}{\partial t}+\boldsymbol{u} \cdot \nabla \boldsymbol{u}\right) &=-\nabla p+\mu \nabla^{2} \boldsymbol{u}+\rho_{f} \boldsymbol{f}_{IB} \label{eq:NS_moment}.
\end{align}
Here, $\boldsymbol{u}$ and $p$ represent the velocity and pressure of the fluid flow, respectively.
The forcing term $\boldsymbol{f}_{IB}$ in (\ref{eq:NS_moment}) represents the immersed boundary force (IB force) to satisfy the non-slip boundary condition on particle surfaces (as elaborated in the following).

To simulate the fluid flow, we numerically solve the incompressible N-S equations (\ref{eq:NS_conti}-\ref{eq:NS_moment}) with a second-order finite difference method \citep{kimImplicitVelocityDecoupling2002}. The temporal advancement from the $n$-th to the $(n+1)$-th time step using the Crank-Nicolson scheme is \citep{kimImplicitVelocityDecoupling2002}:
\begin{align}
    \frac{\boldsymbol{u}^{*}-\boldsymbol{u}^{n}}{\Delta t}+\frac{1}{2}(H( \boldsymbol{u}^{*})+H( \boldsymbol{u}^{n})) &= -G p^{n-1/2}+\frac{1}{2 R e}\left(L \boldsymbol{u}^{*}+L \boldsymbol{u}^{n}\right), \label{eq:NS_first_predict}\\
    \boldsymbol{u}^{**} &= \boldsymbol{u}^* + \Delta t \boldsymbol{f}_{IB}^{n + 1/2}, \label{eq:IB_second}\\
    D G \delta p & = D \boldsymbol{u}^{* *}/\Delta t , \label{eq:pressure_poiss}\\
    p^{n+1/2} & = p^{n-1/2}+\delta p , \label{eq:pressure_update}\\
    \boldsymbol{u}^{n+1} & = \boldsymbol{u}^{* *}-\Delta t G \delta p. \label{eq:vel_update}
\end{align}
Here, $H$, $G$, $L$ and $D$ represent the spatial discrete convection, gradient, Laplacian and divergence operators, respectively, which are calculated by the second-order central-difference scheme on a staggered Eulerian grid \citep{kimImplicitVelocityDecoupling2002}. 
In the present simulations, the computational domain is discretized by a uniform Eulerian grid with the grid spacing of $\Delta h=\Delta x=\Delta y=\Delta z$. 
In (\ref{eq:NS_first_predict}), the first prediction velocity $\boldsymbol{u}^*$ is updated without the consideration of IB forces. Specifically, the approximate factorization of the coefficient matrix through the block LU decomposition is conducted to decouple the different velocity components, and then $\boldsymbol{u}^*$ is obtained by solving a serious of tri-diagonal linear equations without iteration \citep{kimImplicitVelocityDecoupling2002}.
Then, the second prediction velocity $\boldsymbol{u}^{**}$ is calculated via (\ref{eq:IB_second}) with the inclusion of IB forces, as is outlined in equations (\ref{eq:IB_inter}) - (\ref{eq:IB_spread}). Finally, the velocity projection step is performed, including the solution of the Poisson equation (\ref{eq:pressure_poiss}) (in which $\delta p$ is the pressure increment), and the update of pressure and fluid velocity to a new time step via (\ref{eq:pressure_update}) and (\ref{eq:vel_update}). 
To numerically solve the Poisson equation \eqref{eq:pressure_poiss}, we conduct two-dimensional fast Fourier transformations along $x$ and $z$ directions and solve the uncoupled tri-diagonal linear equations along $y$ axis \citep{kimImplicitVelocityDecoupling2002}.
The parameter $Re=U_0L_0/\nu$ presented in equation (\ref{eq:NS_first_predict}) is the Reynolds number based on the characteristic velocity ($U_0$) and length ($L_0$) to normalize the N-S equations (\ref{eq:NS_conti}-\ref{eq:NS_moment}). 
In the present work, we use the settling velocity of an isolated spheroid as the characteristic velocity (i.e. $U_0=V_t$), and the equivalent diameter of the spheroid as the characteristic length scale (i.e. $L_0=D_{eq}$).

Then we move on to the implementation of the IBM. In this method, one needs to allocate a set of Lagrangian marker points to represent the surface of a particle. To do so, we adopt the method proposed by \citet{eshghinejadfard_direct-forcing_2016} to allocate $N_L=2263$ points on the surface of each prolate spheroid. 
As regards the calculation of IB forces presented in equation (\ref{eq:IB_second}), we employ the direct-forcing IBM as follows \citep{uhlmannImmersedBoundaryMethod2005, breugemSecondorderAccurateImmersed2012}:
\begin{align} 
    \boldsymbol{U}^{*}_{l}&=\sum\limits_{ijk} {{\boldsymbol{u}}_{ijk}^{*}\delta \left( {\boldsymbol{X}^{n}_{l}}-{{\boldsymbol{x}}_{ijk}} \right)\Delta {h^3}} , \label{eq:IB_inter}\\
    {\boldsymbol{F}}_{l}^{n+1/2} &= \frac{{{{\boldsymbol{U}_p}(\boldsymbol{X}^{n}_{l})} - {\boldsymbol{U}}_{l}^{*}}}{{\Delta t}} , \label{eq:IB_force}\\
    {\boldsymbol{f}}_{IB,ijk}^{n + 1/2} &= \sum\limits_{l} {{\boldsymbol{F}}_{l}^{n+/2}\delta \left( {{{\boldsymbol{x}}_{ijk}} - {\boldsymbol{X}^{n}_{l}}} \right)\Delta {V_{l}}}. \label{eq:IB_spread}
\end{align}
Note that the above steps should be conducted in synchronization with the flow simulation between equations (\ref{eq:NS_first_predict}) and (\ref{eq:IB_second}).
In the above equations, the capital letters refer to the variables defined on the Lagrangian marker point. 
In equation (\ref{eq:IB_inter}), the first prediction velocity $\boldsymbol{u}^*$ is interpolated from the Eulerian grid to the Lagrangian marker point using the Dirac-delta function $\delta(\cdot)$, in which $\boldsymbol{X}^n_l$ denotes the position of the $l$-th Lagrangian marker point on the particle surface. 
Then, the IB force is calculated on the Lagrangian marker point through equation (\ref{eq:IB_force}), where $\boldsymbol{U}_p$ represents the rigid velocity of the particle. Finally, the IB forces are spread onto the Eulerian grid with the Dirac-delta function via equation (\ref{eq:IB_spread}), in which $\Delta V_l$ represents the volume of the $l$-th Lagrangian marker point. 
The Dirac-delta function used for the transformation of variables on the Eulerian grid and the Lagrangian point is defined by:
\begin{equation} \label{eq:Dirac-Delta}
    \delta \left( {\boldsymbol{x}} \right) = \frac{1}{{\Delta {h^3}}} \cdot \Theta (\frac{x}{{\Delta h}}) \cdot \Theta (\frac{y}{{\Delta h}}) \cdot \Theta (\frac{z}{{\Delta h}}),
\end{equation}
where $\Theta(\cdot)$ is a 3-grid-width discrete Dirac-delta function as \citep{romaAdaptiveVersionImmersed1999}:
\begin{equation} \label{eq:delta}
    \Theta(r)=\begin{cases}\frac{1}{6}(5-3|r|-\sqrt{-3(1-|r|)^2+1)},&\quad0.5\leq|r|\leq1.5\\\frac{1}{3}(1+\sqrt{-3r^2+1}),&\quad|r|\leq0.5\\0,&\quad\text{otherwise}.\end{cases}
\end{equation}

Furthermore, in the implementation of the direct-forcing immersed boundary method, we adopt the multi-forcing scheme with $N_s=2$ iterations to better approximate the non-slip boundary condition on the particle surface \citep{breugemSecondorderAccurateImmersed2012}, and use the inward retraction of Lagrangian marker points with a distance of $r_d=0.3\Delta h$ \citep{breugemSecondorderAccurateImmersed2012} to counteract the excess in the particle effective diameter induced by the finite width of the discrete Dirac-delta function.
One can refer to our previous work \citep{jiangFlowreconstructionBasedApproach2024} for more details about the present numerical method and the validations, in which we simulated the benchmark cases of a single settling sphere or oblate spheroid at different Reynolds numbers \citep{tencateParticleImagingVelocimetry2002, moricheSingleOblateSpheroid2021}.

\subsection{Validation: Drafting-kissing-tumbling of two settling spheres}
\begin{figure}
    \centering
    \includegraphics[width=0.45\textwidth]{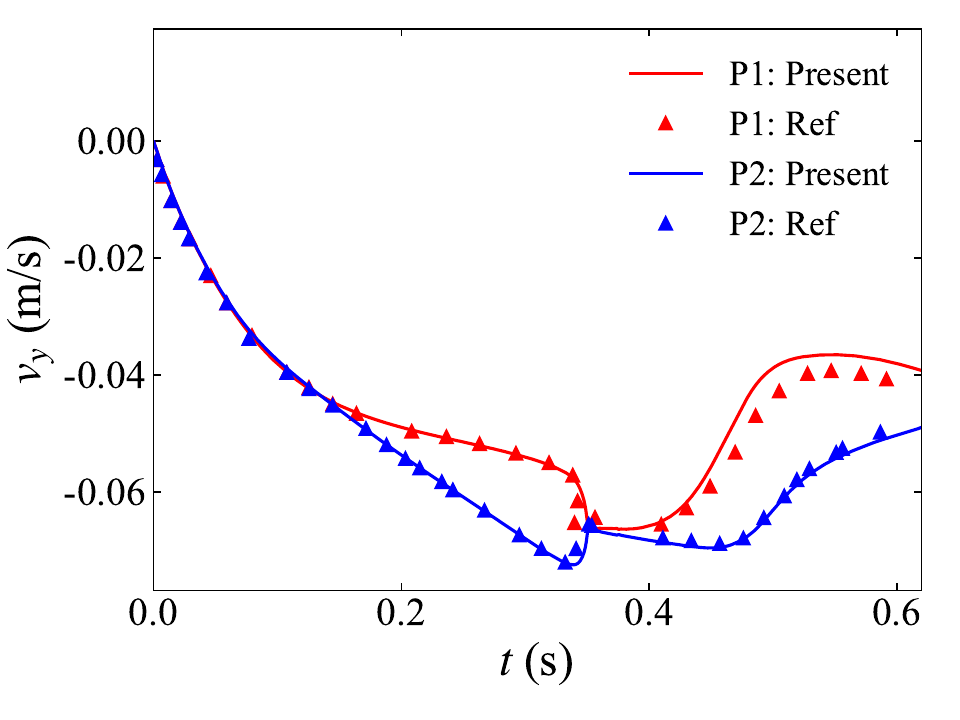}
    \caption{Time evolution of the vertical velocity of two settling spheres undergoing the DKT interaction. P1 and P2 denote the initially leading and trailing particle, respectively. ``Ref" represents the result provided in the reference \citep{breugemSecondorderAccurateImmersed2012}.}
    \label{fig:DKT}
\end{figure}
To further validate the present numerical method in the problems involving particle-particle interactions, we simulate the DKT process of two settling spheres in a closed container \citep{glowinskiFictitiousDomainApproach2001,breugemSecondorderAccurateImmersed2012}. The flow configuration is introduced as follows. The container has a size of $L_x\times L_y \times L_z =[0,1 \mathrm{cm}] \times [0,4 \mathrm{cm}] \times [0,1 \mathrm{cm}]$, and is filled with a Newtonian fluid with a density of $\rho_f=1000 \mathrm{kg/m^3}$ and a kinematic viscosity of $\nu=10^{-6} \mathrm{m^2/s}$. The gravity is applied in the negative $y$ direction with an acceleration of $g=9.8 \mathrm{m/s^2}$. Two spheres with a diameter of $D=0.167 \mathrm{cm}$ and a density of $\rho_p=1140 \mathrm{kg/m^3}$ settle from rest in the container. The initial positions of the two particles are $\boldsymbol{x}_1=(0.495\mathrm{cm},3.16\mathrm{cm},0.495\mathrm{cm})$ (initially leading particle) and $\boldsymbol{x}_2=(0.505\mathrm{cm},3.5\mathrm{cm},0.505\mathrm{cm})$ (initially trailing particle). In the simulation, the computational domain (which is the same as the container) is discretized by $N_x \times N_y \times N_z =96 \times 384 \times 96$ Eulerian grid cells, and the surface of each sphere is represented by $N_L=731$ Lagrangian marker points. Figure \ref{fig:DKT} depicts the temporal evolution of the vertical velocity of the two spheres during the DKT process. We observe that the initially trailing particle is accelerated and settles faster than the leading particle from $t\approx0.15 \mathrm{s}$, and progressively approaches the leading one (drafting stage). At around $t\approx0.34 \mathrm{s}$, the two particles get in touch (kissing stage) and then separate (tumbling stage). 
As shown in figure \ref{fig:DKT}, the velocities of two particles calculated by the present simulation are in agreement with the reference data \citep{breugemSecondorderAccurateImmersed2012} in the drafting stage. While, there is a slight discrepancy between the present result and the reference data after the collision of the two spheres, which is attributed to the difference in the collision model adopted here and in \citet{breugemSecondorderAccurateImmersed2012}.

\subsection{Grid-independence test: sedimentation of an isolated prolate particle} \label{appsec:isolated}
\begin{figure}
    \centering
    \includegraphics[width=0.99\textwidth]{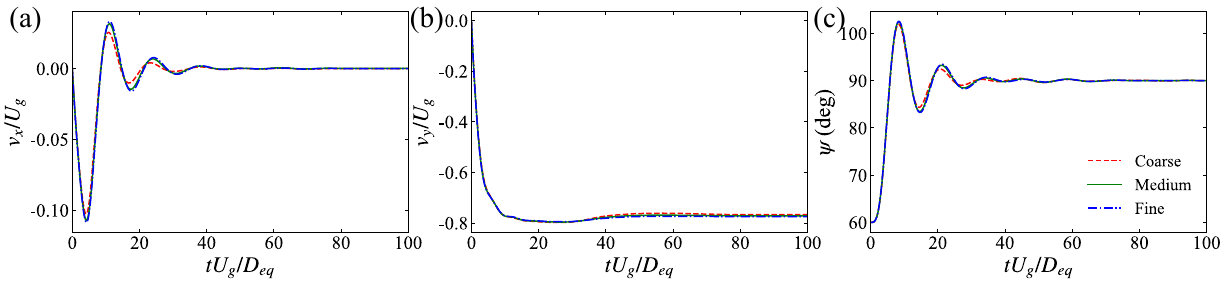}
    \caption{ Time evolution of the (a) horizontal velocity, (b) vertical velocity and (c) pitch angle of the isolated prolate spheroid settling in an initially quiescent fluid. The characteristic velocity $U_g=\sqrt{(\alpha-1)gD_{eq}}$ is used for the normalization.}
    \label{fig:single}
\end{figure}
In this section, we simulate the settling motion of an isolated prolate particle in the quiescent fluid, and examine the influence of grid resolution on the simulation results. The prolate particle with the same parameters as in the main text (i.e. $\lambda=3$, $Ga=80$, $\alpha=2$) is considered. 
In this simulation, we utilize the strategy of moving domain \citep{chen_simulations_2019} to capture the entire settling process of the particle, from rest to steady state, using a computation domain with the size of $L_x \times L_y \times L_z = 12D_{eq} \times 24D_{eq} \times 12D_{eq}$. 
We impose a Dirichlet boundary condition with zero velocity on the bottom boundary, a Neumann boundary condition on the upper boundary of the computational domain, and the periodic boundary condition in the lateral directions. 
At the beginning of the simulation, the prolate spheroid is released from rest with an initial pitch angle of $\psi_0=60^\circ$ ($\boldsymbol{n}_0=(n_x,n_y,n_z)=(\sqrt{3}/2,0.5,0)$). 
To test the grid-independence, three different grid resolutions, i.e. $\Delta h_{coarse}=D_{eq}/16$, $\Delta h_{medium}=D_{eq}/24$ and $\Delta h_{fine}=D_{eq}/32$, are used for the simulation. As shown in figure \ref{fig:single}, the simulation results change a little with the refinement of the grid from $\Delta h =D_{eq}/16$ to $\Delta h =D_{eq}/24$, but the discrepancy between the data obtained by the intermediate-grid and fine-grid simulations is negligible. Therefore, the resolution of $\Delta h =D_{eq}/24$ is sufficient to resolve the fluid-particle interaction under the present parameter setup, and is thus adopted in the simulations in the main text. 
Moreover, figure \ref{fig:single}(c) shows that the prolate spheroid re-orients to the broad-side-on alignment with $\psi=90^\circ$ as the steady orientation under the effect of fluid inertia. The terminal settling velocity is $V_t=0.772 U_g$, yielding a Reynolds number of $Re_t=V_t D_{eq} /\nu =61.8$. This Reynolds number is not high enough to trigger wake instability, so the prolate spheroid settles vertically without oscillation.

\bibliographystyle{jfm}
\bibliography{jfm}

\end{document}